\documentclass[prd,preprintnumbers,superscriptaddress,nofootinbib,twocolumn,aps,10pt]{revtex4-2}
\usepackage{epsfig}
\usepackage{amsmath}
\usepackage{amsfonts}
\usepackage{float}
\usepackage{amssymb}
\usepackage{slashed}
\usepackage[dvipsnames]{xcolor}
\usepackage{pbox}
\usepackage{subfigure}
\usepackage[colorlinks,citecolor=blue, linkcolor = blue, urlcolor = blue]{hyperref}
\usepackage{tabularx}
\usepackage{titlesec}
\usepackage{scrextend}
\usepackage{enumitem}
\usepackage[normalem]{ulem}
\usepackage{orcidlink}
\usepackage{natbib}

\def\gagg{g_{a \gamma \gamma}}
\def\gae{g_{aee}}

\def\bal#1\eal{\begin{align}#1\end{align}}
\def\bea#1\eea{\begin{eqnarray}#1\end{eqnarray}}
\def\beq#1\eeq{\begin{equation}#1\end{equation}}

\def\nn{\nonumber}

\begin{document}


\title{Shedding \textit{Stray Light} on Decaying Light Dark Matter:\\ Constraints from NuSTAR X-ray Observations}

\author{Sk Jeesun\orcidlink{0009-0005-2344-9286}}
 \email{jeesun@sjtu.edu.cn}
 \affiliation{State Key Laboratory of Dark Matter Physics,\\ Tsung-Dao Lee Institute \& School of Physics
and Astronomy, Shanghai Jiao Tong University, Shanghai 200240, China}
\affiliation{Key Laboratory for Particle Astrophysics and Cosmology (MOE) \& Shanghai Key Laboratory for Particle Physics and Cosmology, Shanghai Jiao Tong University, Shanghai 200240,
China}

\author{Tanmoy Kumar\orcidlink{0000-0001-9775-6645}}
 \email{kumartanmoy1998@gmail.com}
 \affiliation{School of Physical Sciences, Indian Association for the Cultivation of Science, 2A \& 2B Raja S.C. Mullick Road, Jadavpur, Kolkata 700032, India}

\begin{abstract}
Light dark matter (DM) (mass $\lesssim \mathcal{O}(100)$ keV) remains challenging to detect in several ongoing indirect detection experiments due to threshold limitations. 
Recent observations of diffuse X-ray photons from the NuSTAR stray-light (SL) data provide a powerful avenue to probe such light DM through its decay signatures in the galactic halo.
This work explores the indirect detection prospects of decaying electrophilic scalar DM, electrophilic and photophilic ALP DM, and dark photon DM using the recent NuSTAR SL data. 
We find that for DM scenarios producing monochromatic two-photon signals, NuSTAR SL data can yield the strongest indirect detection bound in the $7-36$ keV mass range.
In contrast, for dark photon (vector) DM featuring a continuous three-photon spectrum, the strongest indirect detection upper bound arises in the $ 22-65$ keV mass range.
Additionally, we discuss the detection prospects of inelastic DM where the heavier DM decays to a two or three-photon final state along with a massive lighter dark sector particle.
By comparing the resulting continuous photon spectra with the NuSTAR SL data, we obtain the most stringent lower bound on the lifetime of such DM for the mass splitting $\Delta m$ in the range $3 ~{\rm keV}- 100$ keV.

\end{abstract}

\maketitle

\section{Introduction}
A wide range of astrophysical and cosmological observations at different scales provide compelling evidence for the existence of dark matter (DM), a non-luminous and non-baryonic component that constitutes nearly $85\%$ of the matter content of the Universe \cite{Zwicky:1933gu,Rubin:1970zza,Clowe:2006eq,Planck:2018vyg}. 
The existence of cold dark matter is also crucial from the perspective of large-scale structure formation \cite{Planck:2018vyg}.
Despite its overwhelming abundance, the particle nature of DM remains one of the most pressing open questions in particle physics. The absence of any viable dark matter candidate within the Standard Model strongly points towards the existence of new physics beyond it, motivating extensive theoretical and experimental efforts to uncover its origin and properties \cite{Arcadi:2017kky,Roszkowski:2017nbc,Cirelli:2024ssz}.
Weakly interacting massive particle (WIMP) has been the most studied DM model featuring a thermal DM with mass GeV-TeV and interaction strength of the order of weak interactions \cite{Scherrer:1985zt,Srednicki:1988ce}.
The possible connection between DM and the visible sector to achieve the observed relic density  in the early Universe has motivated the community to trace the same interaction in the present day to unravel the particle DM.
Over the past decade, the dark matter (DM) sector has been extensively investigated through a complementary trio of approaches: collider searches \cite{ATLAS:2017bfj,CMS:2017zts, 
Kahlhoefer:2017dnp}, direct detection \cite{Misiaszek:2023sxe,Billard:2021uyg,XENON:2022ltv,XENON:2020rca} and indirect detection \cite{Gaskins:2016cha,PerezdelosHeros:2020qyt}.

Among these, indirect detection of dark matter is particularly compelling, as it looks for observable signals arising from DM pair annihilation or decay to SM particles in the galaxy or dense astrophysical objects \cite{Gaskins:2016cha,PerezdelosHeros:2020qyt}. 
The usual methodology of indirect search is to look for any possible excess in the cosmic ray (electron, proton, quarks, photon or neutrino) over the expected background, often estimated through some power law scaling, and the null observation of BSM signal leads to constraints on the respective interaction channel.
 Thus, indirect detection offers a unique window into the very interactions that shaped the dark matter relic abundance in the early Universe. 
EW scale DM annihilating/decaying to heavy SM quarks (e.g. $b\bar{b}$), charged leptons, vector bosons or high energy photons are already constrained from the observations of telescopes like \textit{Fermi}-LAT \cite{Fermi-LAT:2016afa}, AMS \cite{Bergstrom:2013jra,AMS:2019iwo},  HESS \cite{HESS:2022ygk}.
Indirect searches of DM with high-energy neutrino final states have also constrained such scenarios using the multi-ton neutrino detector like Super Kamiokande \cite{Super-Kamiokande:2015qek}, IceCube \cite{Baur:2019jwm} etc.

Despite heroic efforts to detect dark matter, the persistent null results from both direct and indirect searches have cornered the electroweak-scale DM \cite{Cirelli:2024ssz}.
As a consequence, the keV-sub GeV mass range of DM has drawn great attention from both theorists and experimentalists in recent years~\cite{Nguyen:2024kwy,DelaTorreLuque:2025zjt,Nguyen:2025tkl}.
Particularly for indirect detection, the number density of DM becomes higher for lighter masses, thereby enhancing the possibility of any BSM signal. However, searches in this regime face a key challenge: the finite detection thresholds of experimental setups, which limit their sensitivity to light DM.
For annihilating or decaying DM with mass in the sub-MeV range, the visible product particles are expected to carry energy of roughly the same order or smaller.
Hence, for such keV-scale DM, the only accessible SM final states are neutrinos and photons in the X-Ray energy range.
This poses a unique challenge for ground-based multi-ton neutrino detectors \cite{Arguelles:2019ouk}, including IceCube \cite{Baur:2019jwm} and also for space based experiment like Fermi-LAT \cite{Fermi-LAT:2016afa}. 
Experiments such as INTEGRAL \cite{Bouchet:2011fn,Cirelli:2020bpc}, COMPTEL~\cite{COMPTEL} offer sensitivity at lower energies, enabling them to probe decaying DM with mass down to $\sim$ MeV; however, the sub-MeV regime remains largely elusive.
This stalemate can be disentangled with telescopes such as the Nuclear
Spectroscopic Telescope Array (NuSTAR) telescope \cite{NuSTAR_tech_desc} with sensitivity down to $\mathcal{O}(10)$ keV energy and large effective area which makes it highly sensitive to photons with sub-MeV energies.
Thus, NuSTAR can serve as an ideal indirect probe for keV-scale DM  
which stands as the key focus of this work.

The {Nuclear Spectroscopic Telescope Array} (NuSTAR) is a space based hard X-ray observatory launched by National Aeronautics and Space Administration (NASA) in 2012 that conducts observations in the energy range $3$--$79$ keV~\cite{NuSTAR_tech_desc}. An important feature of the NuSTAR telescope is the presence of a stray light (SL) aperture, an opening between the different optics and detectors of the telescope that allows unfocused photons, which are several degrees off-axis, to reach the detector. Although initially considered a contaminant, this SL component has proven to be an extremely valuable probe of the diffuse X-ray background. 
In this work, we use the SL data of NuSTAR, collected during its 11 years of operation, to search for line-like and continuum X-ray signals originating from the decay of a light DM (mass $\sim$ few keV).
Previously, the same data have been used to set constraints on annihilating DM~\cite{Zakharov:2025coj} as well as sterile neutrino DM~\cite{Krivonos:2024yvm}.
In this work, we consider a series of well-motivated models in which a keV mass DM has an effective coupling with photons and can thus decay to two or more photons, giving rise to observable signals.
Specifically for the following DM models - electrophilic scalar DM, electrophilic and photophilic ALP DM, dark photon DM, we find that the NuSTAR SL data place the strongest indirect constraint in the keV mass range.
On the other hand, for the inelastic fermion DM model, the DM decays to another dark sector particle (with a smaller mass) along with photons.
Using the NuSTAR stray light data, we derive upper bounds on the lifetime 
of such DM candidates with mass splitting down to $\sim$keV, improving upon existing 
indirect detection bounds. 
The strength of these constraints highlights the potential of NuSTAR 
stray light observations as a compelling probe of light dark matter in this 
mass range, which is the primary focus of this work.

This paper is organized as follows. Sec.\ref{sec:stat} contains the detailed discussion about the expected DM signal and the relevant statistical analysis used to set the limit for a generic DM model. In sec.\ref{sec:model} we discuss the expected DM signal from earlier mentioned specific DM models one by one and display the constraints for each model obtained from our analysis. Finally, we conclude in Sec.\ref{sec:discussions}.

\section{Searching for Dark Matter Decay Signatures in NuSTAR Stray Light Data}
\label{sec:stat}

In this section, we describe our strategy to search for signatures of decaying dark matter in the NuSTAR SL data. We begin by outlining the expected photon signal from dark matter decay, followed by a description of the NuSTAR SL observations and their suitability for probing diffuse X-ray emission. We then discuss the signal and background components along with the statistical framework used.

\subsection{Dark matter decay signal}
Light DM in the keV mass range with couplings to SM particles can generically couple to photons either at tree level or at one-loop level. Such a coupling can result in the decay of DM to two or more photons. Such decays give rise to observable X-ray signatures. Depending on the number of final state particles, the resulting spectrum can be either monochromatic with the photon energy $E_\gamma \simeq m_{\rm DM}/2$ for a two-body final state, or a continuum-like spectrum for a multi-body final state.

Such decays occur both within the Milky Way halo and throughout the Universe, giving rise to Galactic and extragalactic contributions to the observed photon flux. The extragalactic component is approximately isotropic and is redshifted, leading to a smooth continuum spectrum. In contrast, the Galactic contribution is not subject to redshift and thus preserves spectral features such as sharp lines. 
The Galactic signal is determined by the dark matter distribution in the Milky Way and scales linearly with the density along the line of sight. Consequently, the resulting photon flux is spatially extended and approximately isotropic at high Galactic latitudes. Moreover, due to the higher dark matter density in the Galactic halo, the Galactic contribution is typically enhanced relative to the isotropic extragalactic component. This makes it well-suited for searches using wide-field X-ray observations. 

Assuming a fixed profile for the Milky Way DM halo, the expected differential photon flux from DM decay is given by
\begin{equation}
    \dfrac{d \Phi_\gamma}{d E_\gamma} = \dfrac{1}{4 \pi} \dfrac{\Gamma_{\rm DM}}{m_{\rm DM}} \dfrac{d N}{d E_\gamma} \left \langle \dfrac{d\, \mathcal{D}_{\rm DM}}{d\, \Omega}\right \rangle \,\, \text{cm}^{-2}\,\text{s}^{-1}\,\text{sr}^{-1}\,\text{keV}^{-1},
    \label{eq:signal}
\end{equation}
where $\Gamma_{\rm DM}$ is the DM decay width, related to the lifetime as $\tau^{-1}_{\rm DM} = \Gamma_{\rm DM}$, $m_{\rm DM}$ is the DM mass.
$dN/dE_\gamma$ is the photon spectrum per decay, and $\langle {d\, \mathcal{D}_{\rm DM}}/{d\, \Omega} \rangle$ represents the stacked differential dark matter column density for the given profile. In this work we use the value of $\langle {d\, \mathcal{D}_{\rm DM}}/{d\, \Omega} \rangle = 5.32\,\, \text{GeV}\,\text{cm}^{-3}\,\text{kpc}\,\text{sr}^{-1}$ as tabulated in~\cite{Krivonos:2024yvm} considering an Navarro–Frenk–White (NFW) profile as given in~\cite{Cautun:2019eaf}.
It is worth highlighting that the particle physics information of the decaying DM across various DM models is encoded in $dN/dE_\gamma$.

\subsection{Statistical Analysis}

In this work, we utilize the NuSTAR stray light (SL) spectral data in the $3$--$18$ keV energy range, constructed and analyzed in Refs.~\cite{Krivonos:2024yvm, Zakharov:2025coj}, based on $\sim 11$ years of observations with the NuSTAR telescope. 
The SL dataset combines observations from both FPMA and FPMB modules of the NuSTAR telescope. After applying data selection, cleaning, and masking procedures to remove contaminated observations and focused X-ray contributions, the final dataset corresponds to a total effective exposure of $t_{\rm obs} \simeq 234$ Ms. To suppress contamination from Galactic ridge X-ray emission, only observations with Galactic latitude $|b| > 3^\circ$ were retained, resulting in a dataset dominated by diffuse emission.
A key quantity characterizing the sensitivity of the SL observations is the \textit{grasp}, defined as the product of the effective detector area and the solid angle subtended by the field of view. In the SL configuration, the effective area is determined by the geometrical size of the detector, while the field of view extends over several square degrees. After accounting for detector masking and data cleaning, the effective grasp is $\mathcal{G} \simeq 40~\text{cm}^2\,\text{deg}^2$ per module, corresponding to a combined grasp of $\mathcal{G} \simeq 80~\text{cm}^2\,\text{deg}^2$ for FPMA and FPMB~\cite{Krivonos:2024yvm, Zakharov:2025coj}.

The detector counts in the SL mode originate from the two components: (i) a nearly uniform instrumental background, and (ii) the diffuse X-ray photons entering through the open sky region~\cite{Zakharov:2025coj}. The latter is the astrophysical signal of interest.
In Refs.~\cite{Krivonos:2024yvm, Zakharov:2025coj}, these two components have been statistically separated, allowing for the reconstruction of the diffuse X-ray spectrum in the energy range $3$--$18$ keV.
In our analysis, we use the diffuse X-ray spectrum obtained from the stacked SL data for $|b| > 3^\circ$ obtained in Ref.~\cite{Zakharov:2025coj}.
The diffuse X-ray spectrum itself consists of two components: X-rays originating from the Sun and the cosmic X-ray background, as discussed in ~\cite{Krivonos:2024yvm,Zakharov:2025coj}.
A potential signal from galactic DM decay would contribute an additional diffuse component to this spectrum. Given the large angular acceptance of the SL mode, the expected signal from Galactic dark matter decay is spatially extended and approximately uniform across the detector for the selected high-latitude observations. Its spectral shape, either line-like or continuum depending on the decay channel, distinguishes it from the smooth astrophysical background.

In our analysis, we adopt a conservative approach by treating the stacked diffuse X-ray spectrum from the SL data as the background and search for any excess that is consistent with the predicted dark matter signal. 
To derive the constraints on the DM parameters, we use a signal-to-noise ratio (SNR) based test statistic. The stacked diffuse X-ray intensity in the $i$-th energy bin, multiplied with the grasp $\mathcal{G}$, the total observation time $t_{\rm obs}$, and the bin width for which we use a conservative value of $\Delta E$,
is treated as the background $B_i$. For a given DM model, the expected signal counts in the $i$-th bin are given by
\begin{equation}
    S_i = \dfrac{d \Phi_\gamma}{d E_\gamma}(E_i) \times \mathcal{G} \times \mathcal{E}_{\rm tot}(E_i) \times t_{\rm obs} \times \Delta E \,,
\end{equation}
where $\dfrac{d \Phi_\gamma}{d E_\gamma}(E_i)$ is the predicted differential photon flux, given in Eq.~\eqref{eq:signal}, evaluated at the central energy $E_i$ of the $i$-th bin. We use the combined grasp of $\mathcal{G} = 80 \,\, \text{cm}^2 \, \text{deg}^2$ 
and fold it with the energy dependent efficiency of the focal plane detectors $\mathcal{E}_{\rm tot}(E_i)$ evaluated at $E_i$, 
an integrated observation time of $t_{\rm obs} = 234$ Ms, and an average conservative bin width of $\Delta E = 0.1$ keV in our analysis~\cite{NuSTAR_tech_desc}.
Following~\cite{Krivonos:2024yvm,Zakharov:2025coj} we have
computed $\mathcal{E}_{\rm tot}(E)$ by directly using the \texttt{DETABS} and the
\texttt{BEABSPAR} calibration files from the NuSTAR Calibration Database
(\texttt{CALDB}). There are two \texttt{DETABS} files for the two focal plane modules (FPMs) FPMA and FPMB, which tabulate the energy-dependent absorption efficiency of the detector surface layer for each of the four CdZnTe detectors on each FPM. We average the \texttt{DETABS} curves over all eight detectors (four each on FPMA and FPMB) to get the $\mathcal{E}_{\rm det}(E)$. The \texttt{BEABSPAR} file, common to both focal plane modules, tabulates the transmission of the Be entrance window as a function of energy $\mathcal{E}_{\rm Be}(E)$. The detector efficiency is then obtained as
\begin{equation}
    \mathcal{E}_{\rm tot}(E) = \mathcal{E}_{\rm det}(E) \times \mathcal{E}_{\rm Be}(E).
\end{equation}
With this the corresponding signal-to-noise ratio is then defined as
\begin{equation}
    \mathrm{SNR} = \sqrt{ \sum_i \frac{S_i^2}{B_i} } \,,
\end{equation}
where the sum runs over all energy bins. 
This statistical method provides a simple and robust estimate of the significance 
of a potential signal excess in the background-dominated regime. We derive 
upper bounds on the DM model parameters by requiring that the predicted 
signal does not exceed $\mathrm{SNR} = 3$, following the Asimov 
approximation of the profile likelihood ratio test statistic \cite{Cowan:2010js}.

\section{Constraints on Decaying Dark Matter}
\label{sec:model}
Having discussed the NuSTAR SL data and the analysis procedure to search for DM decay signals, in this section, we study several well-motivated decaying light (mass $\sim$ few keV) DM models, where the DM either decays to photons directly or via one-loop, and set upper bounds on the DM couplings to SM particles from the NuSTAR SL data.
Throughout the rest of this section, we assume those DM candidates comprise $100\%$ of the total observed DM density (unless mentioned otherwise) without delving into the analysis of their production mechanism.
We now introduce such minimal models one by one.

\subsection{Scalar Dark Matter}
\label{sec:scal}

A spin-$0$ scalar particle is one of the simplest and well-motivated DM candidates that appears in minimal extensions of the SM \cite{Knapen:2017xzo,Bickendorf:2022buy, Montefalcone:2025nmm}.
Such DM particles are also well explored from the perspective of their detectability in current direct and indirect search experiments \cite{Knapen:2017xzo,Mitridate:2021ctr,Ferreira:2022egk}.
Scalar DM can have a variety of couplings with SM particles. Here, we consider a real scalar DM coupled only to electrons and described by the effective Lagrangian,
\begin{eqnarray}
    \mathcal{L}_{\phi}\supset \frac{1}{2}(\partial^\mu \phi)(\partial_\mu \phi)-\frac{1}{2} m_\phi^2 \phi^2-e^\prime_e \phi\,\bar{e}\,e,
\end{eqnarray}
where $m_\phi$ and $e^\prime_e$ are the DM mass and the effective coupling with electron ($e$), respectively.
Such coupling can be generated easily in an Ultraviolet (UV) realization of an extended SM sector featuring 2 scalar doublets and one scalar singlet \cite{Batell:2016ove}. 
However, in this work, we refrain from any kind of UV completion.
Also, throughout this analysis, we remain agnostic about the DM production mechanisms and assume the concerned DM candidate to constitute the total DM relic density.

\begin{figure}
    \centering
    \includegraphics[width=0.9\linewidth]{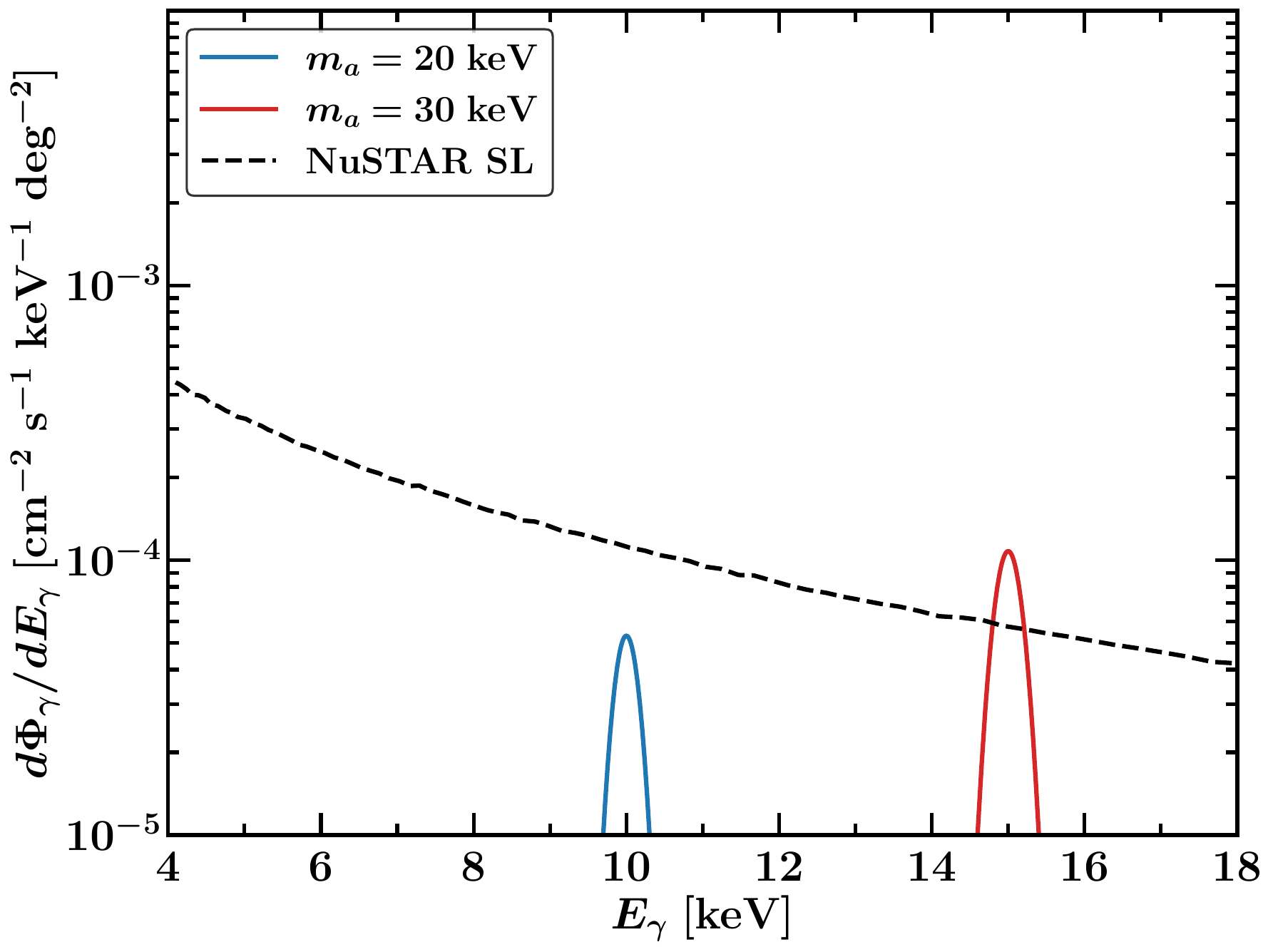}
    \caption{Expected signal photon spectrum from the scalar DM decay as a function of the energy of the emitted photon for $m_\phi=20$ keV (blue solid line) and for $m_\phi=30$ keV  (red solid line) with $e^\prime_e = 10^{-18}$.
    The observed NuSTAR data is shown by the black dashed line.}
    \label{fig:scalar_photon_spectra}
\end{figure}

In this model, although there is no tree level coupling to photons, an effective coupling to two photons can be generated at one-loop through the electron triangle diagram, which can lead to the scalar DM decaying to two photons, thereby producing an observable signal. The loop-induced decay width is given by \cite{Bickendorf:2022buy},
\begin{eqnarray}
    \Gamma_{\phi\to \gamma \gamma}&=&\frac{\alpha_{\rm EM}^2 m_\phi^3}{256\pi^3} \bigg(\frac{e^\prime_e}{m_e}F_{1/2}(x_e) \bigg)^2,
\end{eqnarray}
 where, $\alpha_{\rm EM}$ is the fine structure constant and $x_e=\frac{4 m_e^2}{m_\phi^2}$. 
 In this work, we consider $m_\phi \sim $ few keV $\ll m_e$. Thus in our region of interest $x_l>1$ and $F_{1/2}(x_e)$ is given as~\cite{Bickendorf:2022buy},
 \begin{eqnarray}
     F_{1/2}(x_e)=-2 x_e \bigg[ 1+(1-x_e)\sin^{-2}{\bigg(\frac{1}{\sqrt{x_e}}\bigg)}\bigg].
 \end{eqnarray}

\begin{figure}[t!]
    \centering
    \includegraphics[width=0.95\linewidth]{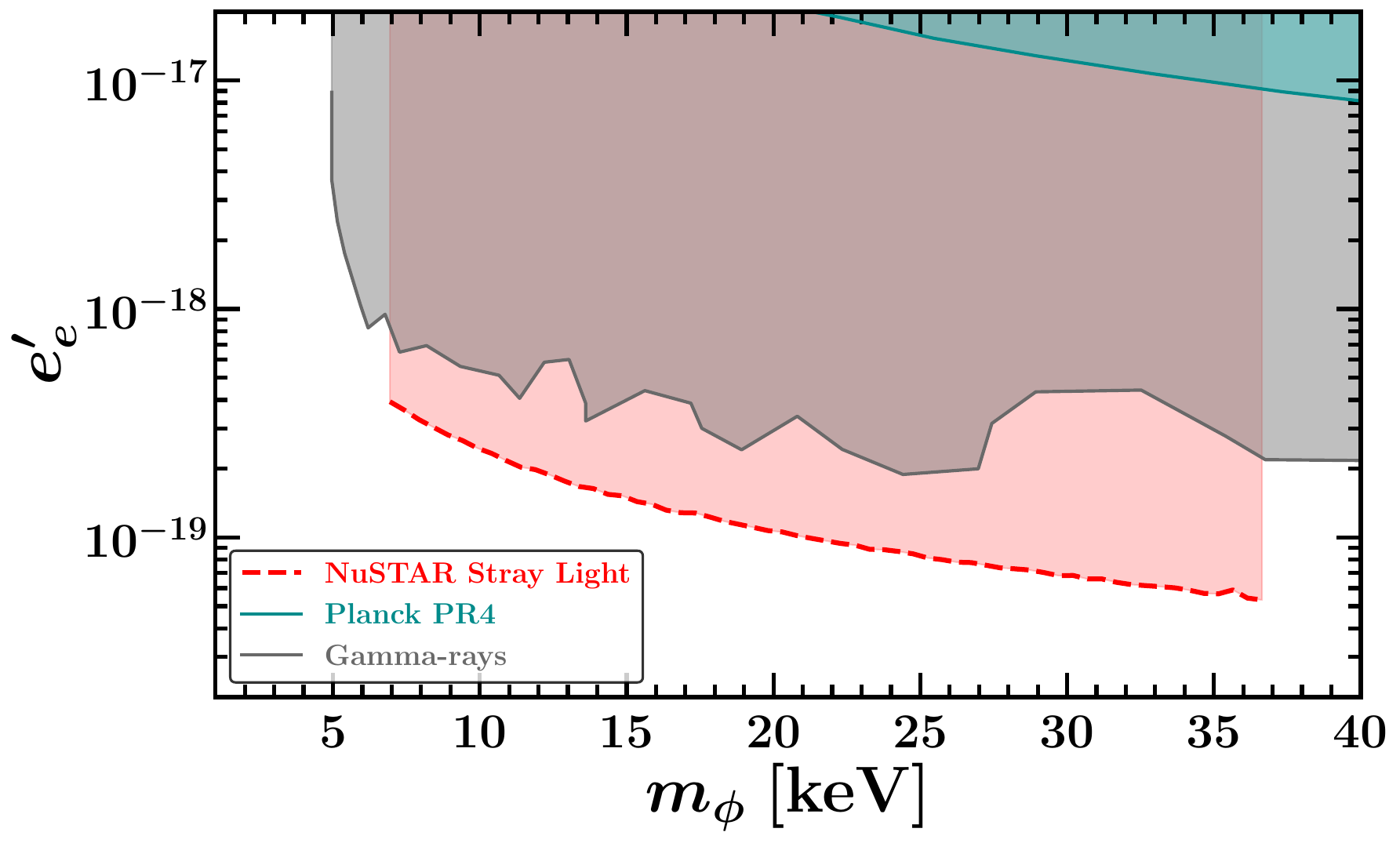}
    \caption{Upper bound on the scalar 
DM-electron coupling $(e^\prime_e)$ as a function of DM mass $m_\phi$ (red dashed) from the NuSTAR SL data, 
derived from loop-induced two-photon decay signatures. Existing constraints from 
other indirect searches~\cite{Foster:2021ngm, Ng:2019gch} and from electromagnetic 
decay of scalar DM using Planck PR4 data~\cite{Montefalcone:2025nmm} are shown by 
the grey and teal shaded regions, respectively.
    }
    \label{fig:scalar-electron}
\end{figure}

Given the decay width, we calculate the expected photon signal using Eq.~\eqref{eq:signal} with $dN/dE_\gamma = 2 \,\delta(E_\gamma - m_\phi/2)$\footnote{ To account for the finite detector resolution, we model the delta function as a Gaussian centered at $m_{\rm DM}/2$ with $\sigma = \text{FWHM}/2 \sqrt{2 \ln{2}}$. For the Full Width at Half Maximum (FWHM) we use a conservative value of $400$ eV~\cite{NuSTAR_tech_desc}.}.
The expected signal photon spectra from the decay of scalar DM to two photons via an electron loop for a fixed coupling $e^\prime_e = 10^{-18}$ are displayed for two benchmark masses: $m_\phi=20$ keV (blue solid line) and for $m_\phi=30$ keV  (red solid line) in Fig.\ref{fig:scalar_photon_spectra}.
Note that both of the decay spectra are peaked at $m_\phi/2$ as expected from the kinematics.
 For a lower value of $m_\phi$, the amplitude of the photon spectra decreases due to the mass suppression of the decay width.
The observed NuSTAR spectra are also shown for comparison, depicted by a black dashed line.

Using this expected photon signal, we compute the upper bound on the DM coupling (to electrons) from the NuSTAR SL data, as a function of DM mass using the procedure mentioned in \ref{sec:stat}, and it is shown in Fig.\ref{fig:scalar-electron} in the $m_\phi - e^\prime_e$ plane by the red dashed line.
As expected, the limit becomes weaker at low $m_\phi$ due to the decreased signal event rate as shown earlier in Fig.\ref{fig:scalar_photon_spectra}.
The sharp decrease in the limit at $m_\phi\sim 6$ keV is due to the fact that the minimum observed energy of the SL photon is $\sim 3$ keV ,which can only arise from a scalar DM of mass $\sim 6$ keV.
In the same plane, we also display existing indirect search limits from Ref.~\cite{Montefalcone:2025nmm}, combining XMM-Newton \cite{Foster:2021ngm} and previous NuSTAR M31 data \cite{Ng:2019gch} shown by the grey shaded region.
CMB constraint on the electromagnetic decay of scalar DM using Planck PR4 data \cite{Montefalcone:2025nmm} has been indicated by the teal blue shaded region.
The indirect search limit from INTEGRAL \cite{Laha:2020ivk} starts to dominate for $m_\phi\gtrsim 100$ keV, and hence, we do not show it here.
On the other hand, astrophysical constraints like stellar cooling limit exclude $e^\prime_e\gtrsim 10^{-15}~{\rm and}~ m_\phi\lesssim10$ keV \cite{Hardy:2016kme}.
Despite all the stringent constraints, we find that NuSTAR SL data can explore previously un-excluded parameter space and provide the strongest limit for $5~{\rm keV}\lesssim m_\phi \lesssim 35$ keV.

\subsection{ALP (Pseudoscalar) Dark Matter}
Axion is a spin-$0$ pseudoscalar particle that originates as a pseudo-Nabu-Goldstone boson (pNGB) of a spontaneously broken global 
$U(1)$ symmetry (also known as Peccei-Quinn (PQ) symmetry) in the Peccei-Quinn solution to the strong CP problem in QCD~\cite{Peccei:1977hh, Weinberg:1977ma, Wilczek:1977pj,Kim:2008hd}.
The axion exhibits a unique feature where its mass is inversely proportional to the $U(1)$ symmetry breaking scale and thus attracts severe constraints on the DM parameter space \cite{DiLuzio:2020wdo, Co:2019jts}.
However, axion-like particle (ALP) represents a generic pNGB realized in various BSM scenarios with $U(1)$ symmetries broken at some scale higher than the electroweak (EW) scale and not limited to the strong CP aspect  \cite{Biekotter:2025fll,DiLuzio:2020wdo,Bharucha:2022lty}. Generically ALPs can have masses ranging from a few eV to several GeV and can have extremely feeble but non-zero couplings to SM quarks, leptons as well as gauge bosons \cite{Bauer:2017ris,Ferreira:2022egk}. Consequently, ALPs are also one of the most promising and extensively studied DM candidates \cite{Dine:1982ah,Preskill:1982cy, 
Abbott:1982af, DiLuzio:2020wdo, Co:2019jts}.
In this section, we analyze the detection prospects of a $\mathcal{O}$(keV) mass ALP DM decaying to two photons using NuSTAR SL data. 

We consider the following two EFT scenarios for the ALP DM-- (1) photophilic ALP DM: ALP has coupling to photon only and (2) electrophilic ALP DM: ALP has coupling to electron only.
We discuss them separately in the following two subsections.

\begin{figure}[!tbh]
    \centering
    \includegraphics[width=0.9\linewidth]{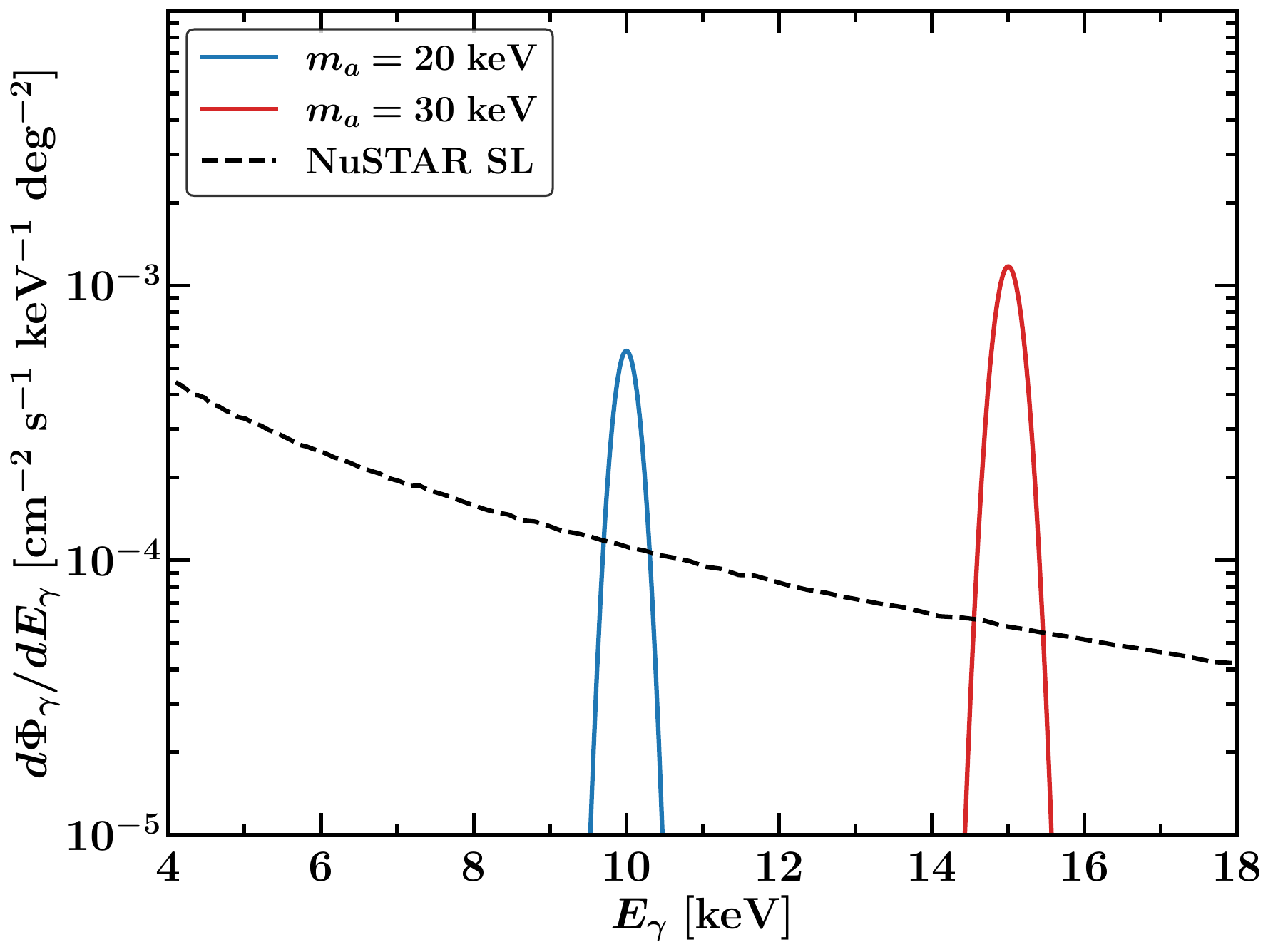}
    \caption{Expected signal photon spectrum from the decay of the ALP DM as a function of the photon energy for $m_a=20$ keV (blue solid line) and for $m_a=30$ keV  (red solid line) with $\gagg = 10^{-17}\, \mathrm{GeV}^{-1}$.
    The observed NuSTAR data is shown by the black dashed line.}
    \label{fig:alp_photon_spectra}
\end{figure}

\subsubsection{ Photophilic ALP Dark Matter}
We consider an ALP ($a$) with mass $m_a$ described by the effective lagrangian,
\begin{equation}
\mathcal{L}_{\rm ALP} = \frac{1}{2} \partial_\mu a \partial^\mu a - \frac{1}{2}m^{2}_{a} a^2 + \dfrac{g_{a\gamma\gamma}}{4} a F^{\mu \nu} \Tilde{F}_{\mu \nu}.
\end{equation}
Here, $F_{\mu \nu}$ corresponds to the photon field tensor and $g_{a\gamma\gamma}$ represents the ALP-photon effective coupling.
We do not assume any other SM coupling of this ALP in the EFT framework. Given the coupling to photons, the ALP can decay to two photons and the corresponding decay width of $a$ is given by,
\begin{eqnarray}
    \Gamma_{a\to\gamma\gamma}= m_a^3\frac{g_{a\gamma\gamma}^2}{64\pi}
    \label{eq:axion_deacy_width}
\end{eqnarray}
with the energy of the produced photon being $E_\gamma = m_a/2$. If $m_a \sim \text{few keV}$, the produced photons from the decay of ALP DM will contribute to the NuSTAR SL data.

\begin{figure}[ht!]
    \centering
    \includegraphics[width=0.95\linewidth]{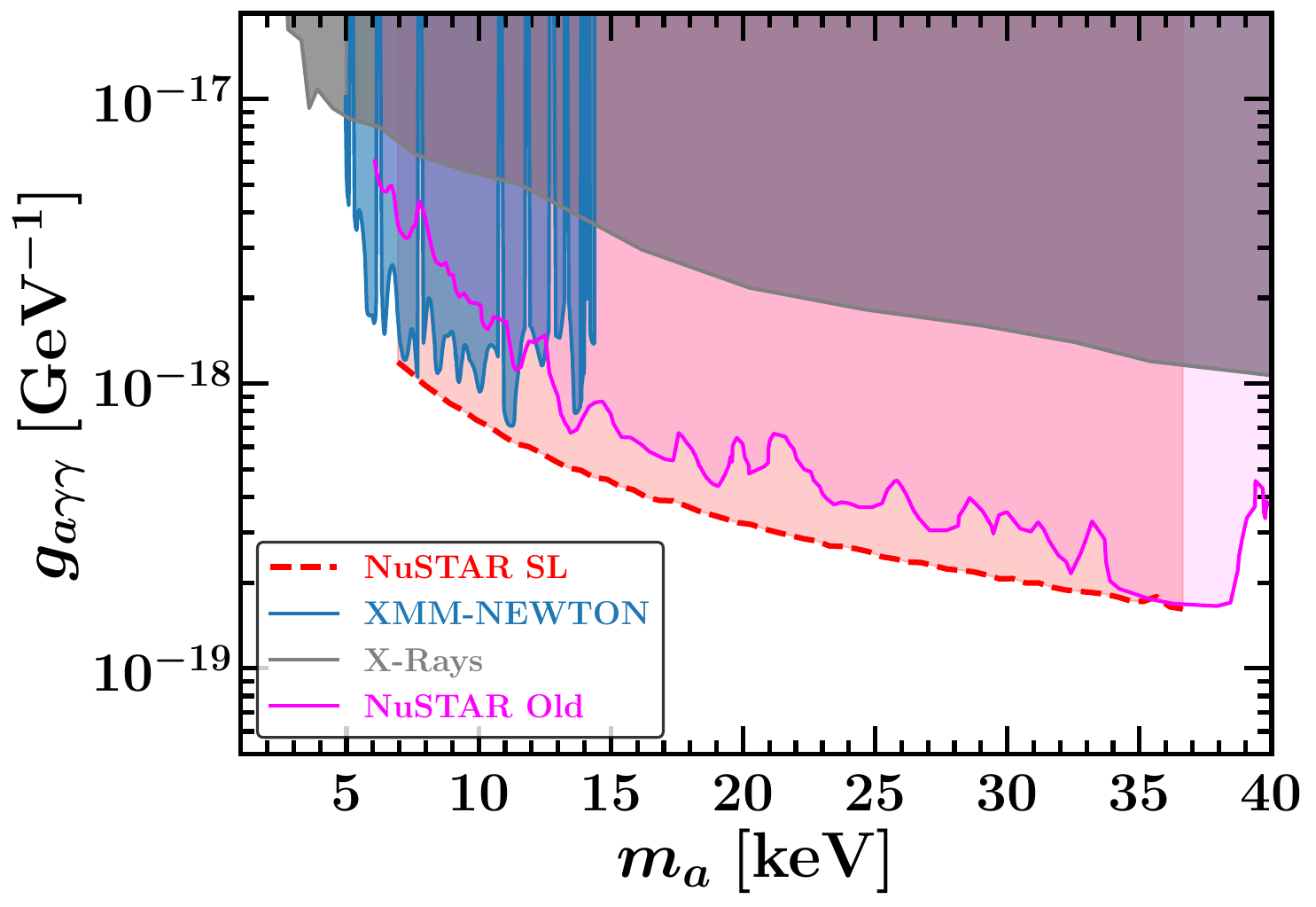}
    \caption{Upper bound on ALP-photon coupling ($\gagg$) from NuSTAR SL data (red dashed line) w.r.t. the DM mass $m_a$. Existing constraints from other indirect searches like XMM-Newton (blue shaded region) \cite{Foster:2021ngm}, previous NuSTAR M31 data (magenta shaded region) \cite{Ng:2019gch} and other X-ray searches (grey shaded region) \cite{Cadamuro:2011fd} are also showcased with different colors.}
    \label{fig:alp-photon}
\end{figure}

Just like the scalar DM case, the differential photon flux is obtained by substituting the decay width of Eq.~\eqref{eq:axion_deacy_width} in Eq.~\eqref{eq:signal} with $dN/dE_\gamma = 2 \,\delta(E_\gamma - m_\phi/2)$. The resulting differential photon flux is shown in Fig.~\ref{fig:alp_photon_spectra} for two different ALP masses i.e. $m_a=20$ keV (blue solid line) and $m_a=30$ keV  (red solid line), with a fixed photon coupling $\gagg = 10^{-17}\,\, \text{GeV}^{-1}$. Also shown is the NuSTAR SL spectra (black dashed line) for comparison. 
The shape of the signal spectra and its change with ALP DM mass can be understood analogously to our previous discussion in the context of scalar DM (see Fig.\ref{fig:scalar_photon_spectra} and its discussion).

Given the photon flux, we follow the procedure discussed in Sec.\ref{sec:stat} and calculate the upper bound from NuSTAR SL data 
on $\gagg$ and it is shown in Fig.\ref{fig:alp-photon} in the $m_a - g_{a\gamma\gamma}$ plane by the red dashed line.
As expected from our previous discussion, the limit on $g_{a\gamma\gamma}$ decreases with a decrease in $m_a$ featuring a sharp cut off at $m_a\sim6$ keV.
For comparison, we also show the existing direct search limits from XMM-Newton (blue shaded region)\cite{Foster:2021ngm}, previous NuSTAR M31 data (magenta shaded region) \cite{Ng:2019gch}, and other X-ray searches (gray shaded region) \cite{Cadamuro:2011fd}.
We find that, in some mass ranges the NuSTAR stray light can provide almost a factor $\sim 2$ stronger limit than existing searches. 
A recent  search using eROSITA observation of the
Large Magellanic Cloud sets the strongest limit on $g_{a\gamma\gamma}$
for $2~\mathrm{keV} \lesssim m_a \lesssim 5~\mathrm{keV}$~\cite{TerolCalvo:2026nlr} complementary to our region of interest.

\subsubsection{Electrophilic ALP Dark Matter}

\begin{figure}[!tbh]
    \centering
    \includegraphics[width=\linewidth]{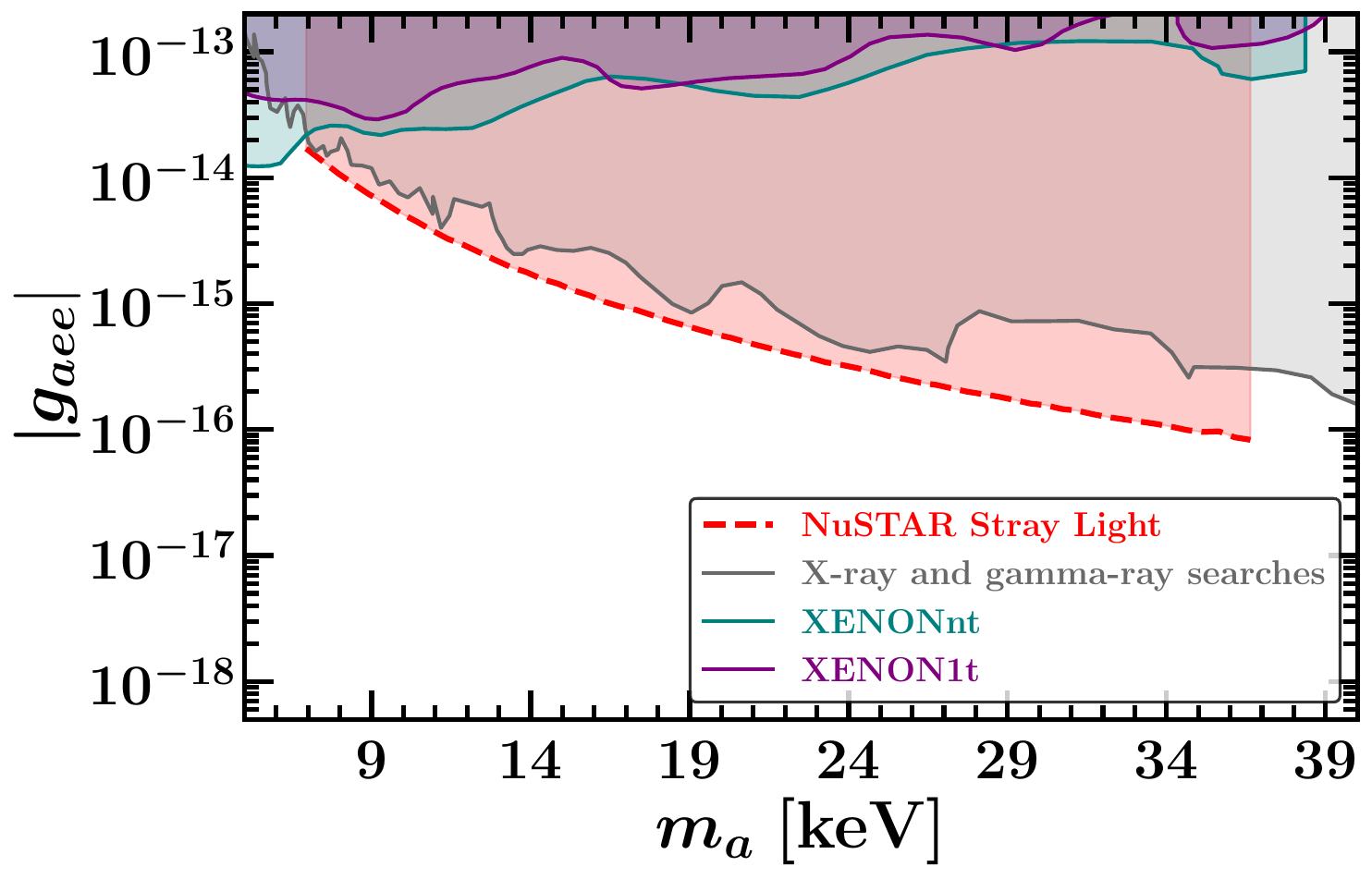}
    \caption{Upper bound on ALP-electron coupling ($g_{aee}$) from NuSTAR SL data (red dashed line) as a function of the DM mass $m_a$. Existing constraints from direct searches like XENONnT (blue shaded region)~\cite{XENON:2022ltv} and XENON1t (purple shaded region)~\cite{XENON:2020rca} and from previous X-ray and gamma-ray searches (gray shaded region) \cite{Ferreira:2022egk} are also shown for comparison.}
    \label{fig:alp-electron}
\end{figure}

Alternatively, another popular scenario is an ALP DM candidate that couples only with electrons given by the following effective Lagrangian~\cite{Pospelov:2008jk},
\begin{equation}
\mathcal{L}_{\rm ALP} = \frac{1}{2} \partial_\mu a \partial^\mu a - \frac{1}{2}m^{2}_{a} a^2 -\frac{g_{aee}}{2 m_e} \partial_\mu a\, \bar{e}\,\gamma^\mu \gamma^5 e ~.
\end{equation}
If such an ALP is light with $m_a<2 m_e$, it can not decay to electrons, thereby seemingly making it challenging to detect.
However, given a coupling with electron only, 
an effective ALP-photon coupling can be generated at one-loop (through $e$ triangle diagram) which is given by \cite{Bauer:2017ris,Ferreira:2022egk},
\begin{eqnarray}
    g_{a\gamma\gamma}^{\rm loop} &=& \frac{\alpha\, g_{aee}}{\pi\ m_e} \left(1-\tau f^2(\tau)\right)~~{\rm where,} \\
    f(\tau)&= &
              \sin^{-1}\left(\frac{1}{\sqrt{\tau}} \right) {\rm for}~ \tau\geq 1,
\end{eqnarray}
with $\tau=4 m_e^2/m_a^2$.
This induced photon coupling can trigger the decay of light electrophilic ALP DM to two photons. 
Thus, the electrophilic ALP DM will eventually lead to a two-photon signature which can enable us to search for it in the same methodology prescribed earlier.

We follow the same procedure as the photophilic case to extract the upper bound on $g_{aee}$ from NuSTAR SL data and it is shown in Fig.\ref{fig:alp-electron} in the $m_a-g_{aee}$ plane.
As existing constraints, we portray the direct search limit from XENON1t and XENONnT (purple and blue shaded regions) \cite{XENON:2020rca,XENON:2022ltv} and from previous X-ray and gamma ray searches (gray shaded region) \cite{Ferreira:2022egk}.
We find that in the keV mass range, the upper bound derived from the NuSTAR SL data can be orders of magnitude stronger than existing direct detection limits and surpasses previous astrophysical constraints on ALP DM by a significant margin.

\subsection{Vector Dark Matter}
\label{sec:vec}
Light vector bosons arise in many well-motivated BSM scenarios \cite{Rizzo:2018ntg,Caputo:2021eaa} and have been extensively studied as cold DM candidate \cite{Pospelov:2008jk,Redondo:2008ec, Nelson:2011sf,Arias:2012az,Farzan:2014foo}. In this section, we analyze the indirect detection prospect of keV-scale vector DM. As an example, we consider the popular dark photon (DP) DM with kinetic mixing with the SM photon. However, our results can be easily translated to other ``electrophilic" vector bosons as well.
We consider the Lagrangian,
\begin{eqnarray}
    \mathcal{L}_{\rm DP} = -\frac{1}{2}m_{A'}^2 A'^\mu A'_\mu -\frac{1}{4}F'^{\mu \nu} F'_{\mu \nu}-\frac{\epsilon}{2} F'^{\mu \nu} F_{\mu \nu},
\end{eqnarray}
where $A'$ is the dark photon with corresponding field strength tensor $F'^{\mu \nu}$ and mass $m_{A'}$.
On the other hand, the usual SM photon field strength tensor is denoted as $F^{\mu \nu}$ and $\epsilon$ is the kinetic mixing strength.
Because of this mixing, DPDM can couple to the SM fermions with an effective coupling $\epsilon\,\alpha_{\rm EM}$,  with $\alpha_{\rm EM}$ being the electromagnetic fine structure constant. If the DPDM is sufficiently heavy, then it can decay to SM fermions, for example electron-positron pair if $m_{A'}>2 m_e$, which can produce observable signals, thus providing a robust indirect signature.

\begin{figure}[h!]
    \centering
    \includegraphics[width=0.9\linewidth]{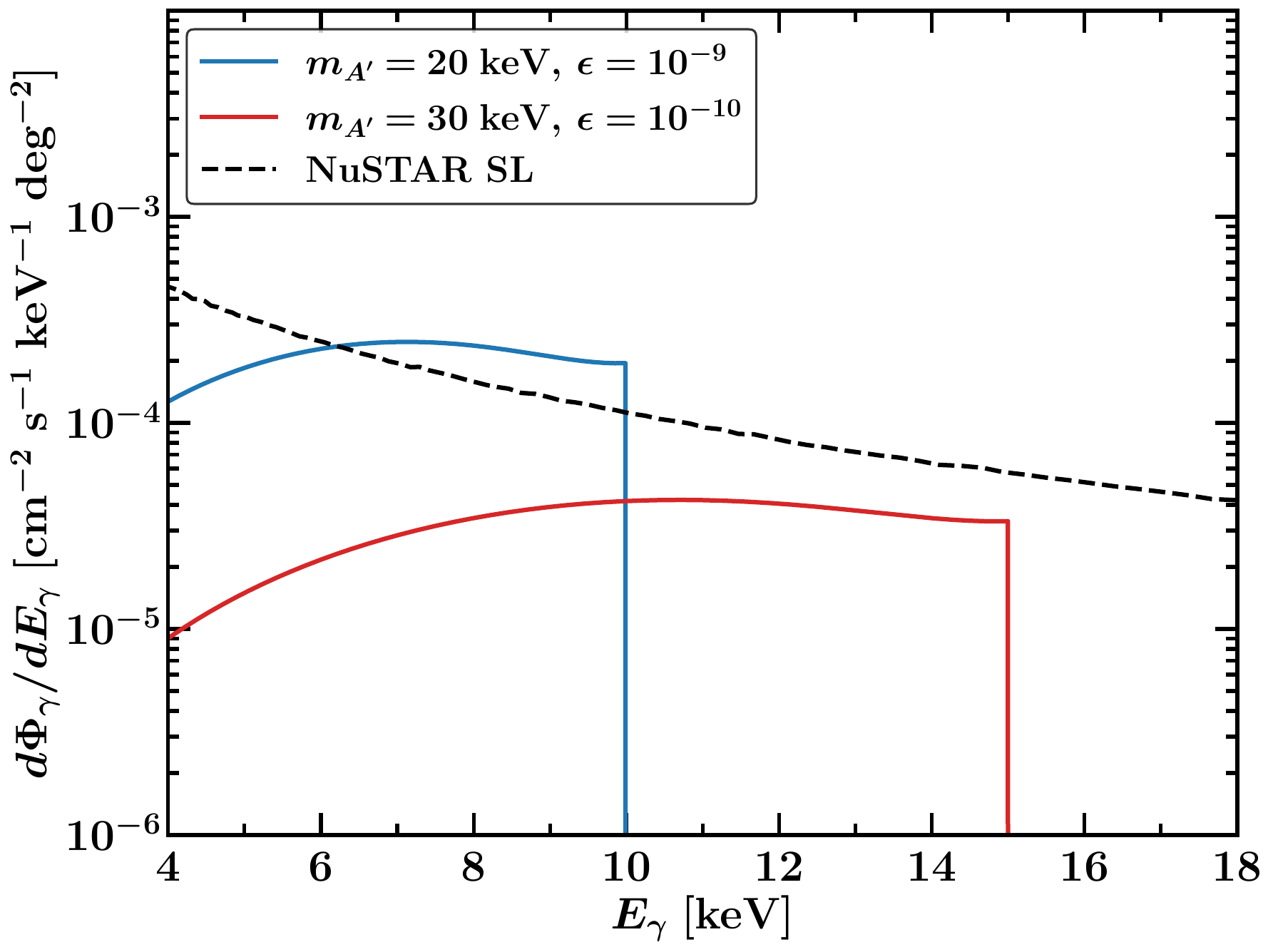}
    \caption{Expected signal photon spectrum from the DPDM trident decay as a function of the emitted photon energy for $m_{A'}=20$ keV with $\epsilon=10^{-9}$ (blue solid line) and for $m_{A'}=30$ keV with $\epsilon=10^{-10}$ (red solid line).
    The observed NuSTAR background is shown by the black dashed line.}
    \label{fig:dp_spec}
\end{figure}

However, if $m_{A'}\ll2 m_e$ then indirect detection of DP DM becomes more complicated since it can not decay to fermions. Furthermore, the decay to two photons via fermion loops is also forbidden due to the spin angular moment argument \cite{Landau:1948kw,Yang:1950rg}.
In this mass regime, the only visible signature possible is the trident decay of DP, $A' \rightarrow 3\,\gamma$, through the quadrilateral loop of SM fermions~\cite{Pospelov:2008jk,Linden:2024fby}.
Using this trident decay, INTEGRAL was able to constrain the kinetic mixing $\epsilon$ for $m_{A'}\gtrsim 60$ keV \cite{Linden:2024fby}.
We demonstrate that NuSTAR, with its operating range spanning $6-30$ keV, can serve as an ideal probe of such DPDM and complement the INTEGRAL bound in the low-mass regime.

The DP DM trident decay width in the limit $m_{A'}\ll 2 m_e$ is given by \cite{Pospelov:2008jk,Linden:2024fby},
\begin{equation}
    \Gamma_{A' \rightarrow \gamma \gamma\gamma} = 1 \,\text{sec}^{-1} \, \times \, \left( \dfrac{\epsilon}{0.003} \right)^2 \, \times \, \left( \dfrac{m_{A'}}{m_e} \right)^9
\end{equation}
and the corresponding spectrum of the produced photons from this decay is given by,
\begin{eqnarray}
   \left( \frac{dN_\gamma}{dE_\gamma}\right)_{A'\to3 \gamma}= \frac{2 x^3}{17 m_{A'}} \bigg(1715-3105 x+\frac{2919}{2}x^2 \bigg),
\end{eqnarray}
where $x= 2E_\gamma/m_{A'}$($0<x<1$) and $E_\gamma$ is the photon energy. Note that the spectrum of the produced photons in the case of the trident decay is significantly different from that of the two-photon decay. The two-photon decay spectrum is a delta function centered at $E_\gamma = m_{\rm DM}/2$ and while the photon spectrum from the trident decay of DP DM exhibits a continuous distribution. 

Given the decay width and the photon spectrum, we calculate the flux of photons from DP DM decays in the galaxy following Eq.~\eqref{eq:signal}, and the resulting flux is shown in Fig.\ref{fig:dp_spec} 
for two different DP DM masses and kinetic mixings, viz., $m_{A'}=20$ keV with $\epsilon=10^{-9}$ (blue solid line) and with $m_{A'}=30$ keV with $\epsilon=10^{-10}$ (red solid line. 
The sharp cut off in the signal spectra at $E_\gamma=m_{A'}/2$ is reflected from the kinematics of the three-photon decay~\cite{Linden:2024fby}.
For comparison, the observed NuSTAR SL background is shown by the black dashed line.

\begin{figure}[t!]
    \centering
        \includegraphics[width=0.95\linewidth]{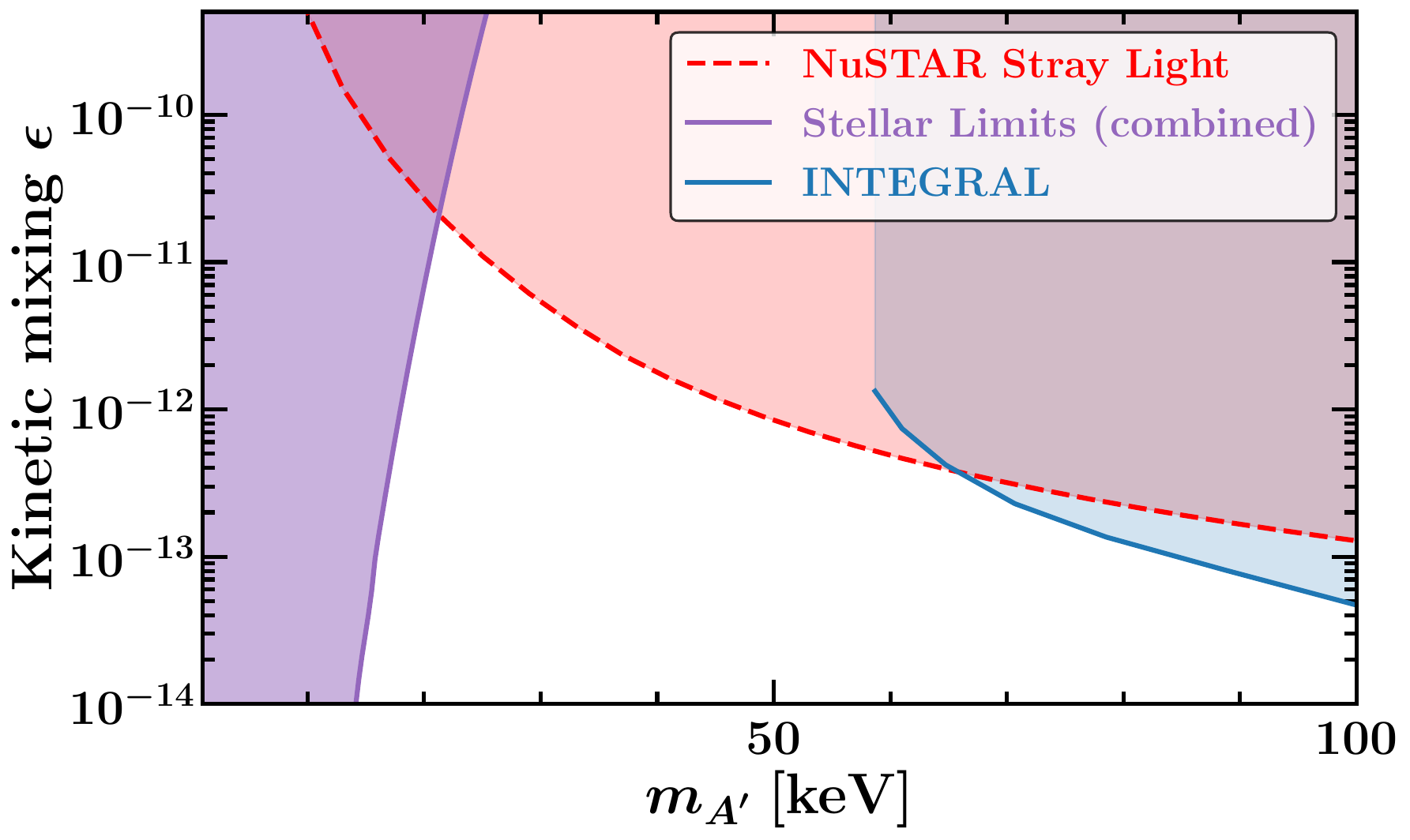}
    \caption{Upper bound on DP kinetic mixing parameter ($\epsilon$) from the NuSTAR SL data as a function the DP DM mass $m_{A'}$ (red dashed line). Existing constraints from indirect searches like INTEGRAL (cyan shaded region) \cite{Linden:2024fby} and combined stellar cooling limits (magenta shaded region) \cite{An:2013yfc,Giannotti:2015kwo} are shown with different colors.}
    \label{fig:photon-darkphoton}
\end{figure}

Having discussed the spectrum, we calculate the upper bound on $\epsilon$ from the NuSTAR SL data following the procedure outlined in Sec.~\ref{sec:stat} and the resulting constraint in the $m_{A'}$ vs. $\epsilon$ plane is shown in Fig.\ref{fig:photon-darkphoton} by the red dashed line.
In contrast to two-photon decays, where the maximum mass that we can probe is equal to two times the energy of the highest energy bin, in case of the three-photon decay signal of DPDM there is no such cutoff due to the continuous nature of the spectrum. 
Hence we find that NuSTAR can probe $m_{A'}\gtrsim36$ keV 
setting new limits in that range. 
The limit improves with a higher DM mass due to the combined effect of less suppression by $m_{A'}$ in the decay width and spectrum, as well as the smaller background at higher energies.
However, for $m_A'\gtrsim 60$ keV, INTEGRAL \cite{Linden:2024fby} provides the most stringent constraint as shown by the cyan shaded region.
Stellar cooling limit (magenta shaded region) \cite{An:2013yfc,Giannotti:2015kwo} becomes dominant for $m_{A'}\lesssim 20$ keV.
The constraint from NuSTAR SL data can be competitive to the direct search limits \cite{XENON:2021qze}, and provides one of the most stringent indirect detection limits in the keV mass range.

\subsection{Inelastic (Fermionic) Dark Matter}
Finally, we consider another well-motivated light DM scenario, which is the inelastic 
DM, where the dark sector comprises multiple BSM particles, thereby enabling rich phenomenology.
In this scenario, one has two fermions $\chi_{1,2}$ with a tiny mass splitting $\Delta m$ between them, and together they constitute the total DM abundance.
The striking feature of this setup is that for sufficiently small couplings with the SM, the heavier state becomes a long-lived particle. However, at later times, it can decay to the lighter state, accompanied by one or more SM particles, which can be probed in different indirect detection experiments.
If the mass splitting $\Delta m\lesssim 2 m_e$, the only visible final state can be a photon, which can be searched for using existing and future telescopes, thereby setting new limits on the parameters of this model.  
Here, we use the NuSTAR SL data to probe such inelastic decaying DM.

\begin{figure}[!tbh]
    \centering
    \includegraphics[width=0.9\linewidth]{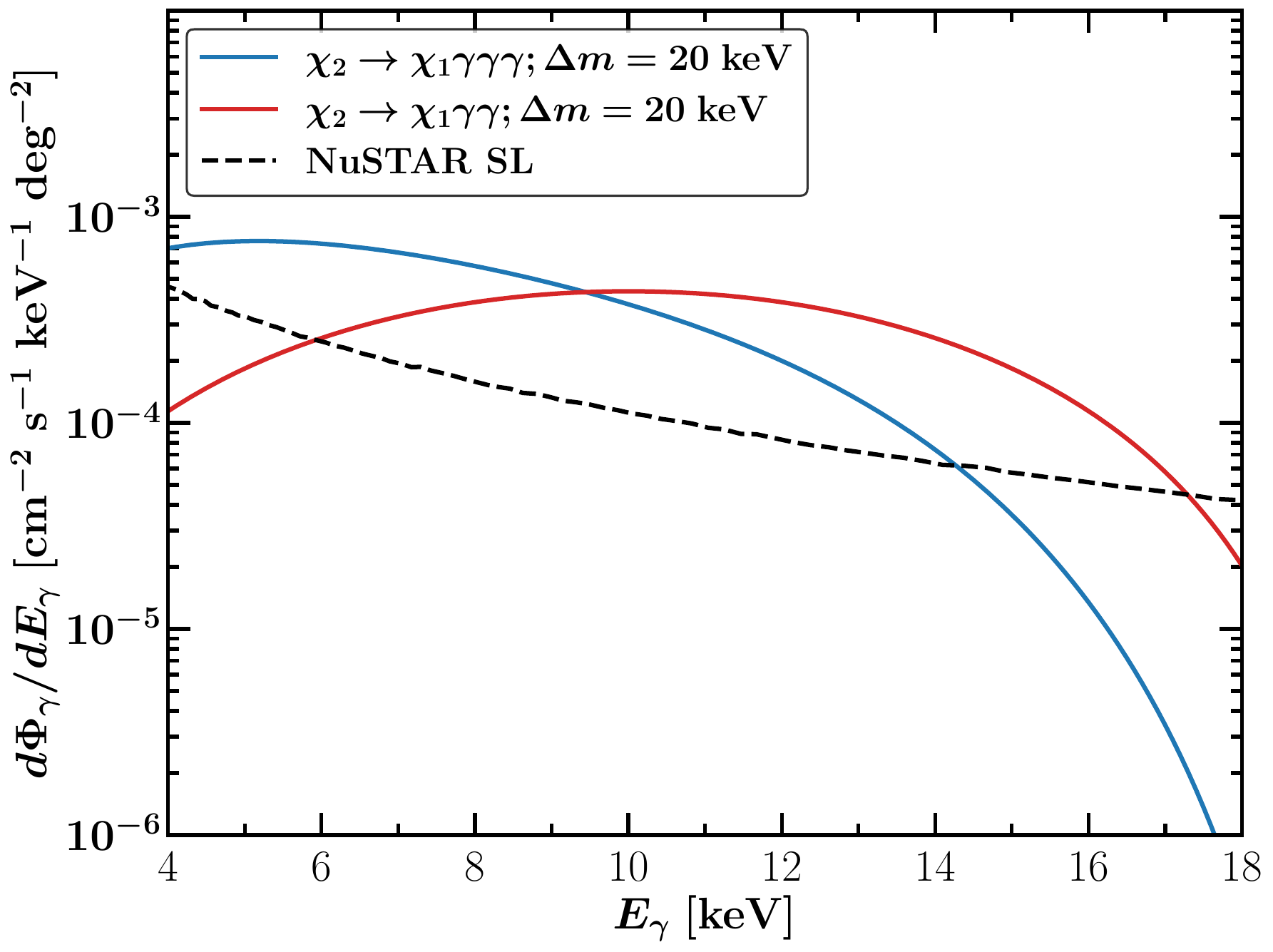}
    \caption{Expected signal photon spectrum from the decay $\chi_2\to \chi_1 \gamma \gamma\gamma$ (blue solid line) and $\chi_2\to \chi_1 \gamma \gamma$ (red solid line) as function of the emitted photon energy for $\Delta m=20$ keV with $\tau_{\chi_2} m_\chi = 10^{20}$.
    The observed NuSTAR spectra are shown by black dashed line.}
    \label{fig:spec_inela}
\end{figure}

We consider two case - \textit{Case 1}: DM $\chi_{1,2}$ couples in-elastically to two photons through the effective operator, $\chi_1 \chi_2 F_{\mu \nu}F^{\mu \nu}$,
where $\Lambda$ is some scale. We assume $\chi_2$ to be the heavier state, $m_{\chi_2}>m_{\chi_1}$, for which $\chi_2$ can decay through the process $\chi_2\to \chi_1+2\,\gamma$, where $m_i$ denotes the mass of $\chi_i~(i=1,2)$.
The mass splitting in this case is given by $m_{\chi_2}-m_{\chi_1}=\Delta m$.
One scenario could be $\chi_2,\chi_1$ are two Majorana fermions which couple to the SM fermion through a scalar portal. As mentioned in Sec.\ref{sec:scal}, the scalar can couple to 2 photons through a fermion loop, leading to the above effective operator~\cite{Krnjaic:2025zjl}.
We also consider another scenario --\textit{Case 2:} The Majorana fermion DM ($\chi_2,\chi_1$) couples (inelastically) to a massive dark photon mediator, where the following process $\chi_2\to \chi_1+A'$ can be realized. However, for $\Delta m\lesssim m_A'<2 m_e$, the above decay is forbidden and the only visible signal can be through the trident decay of dark photon to three SM photons as mentioned previously in Sec.~\ref{sec:vec}. In such scenario the decay $\chi_2\to \chi_1+3\,\gamma$ can be realized \cite{Krnjaic:2025zjl}. Note that in either of the scenarios, apart from the photons, there also exists a massive (invisible) final state  particle. As a consequence, in contrast to the popular DM decays with massless particle (e.g. sterile neutrino DM \cite{Krivonos:2024yvm}), the inelastic DM decays lead to a quite different spectrum of the signal photon, as will be discussed next.

In what follows, we keep our analysis generic and constrain the DM lifetime vs. the DM mass. However, given a UV complete model featuring such interactions, our bounds can be easily translated to bounds on the model parameters. The differential photon flux from inelastic DM decay in the galaxy is given by Eq.~\eqref{eq:signal} with $m_{\rm DM} = \Delta m$, $\Gamma_{\rm DM} = 1/\tau_{\chi_2}$ and $\langle {d\, \mathcal{D}_{\rm DM}}/{d\, \Omega} \rangle$ being replaced with $1/2 \, \langle {d\, \mathcal{D}_{\rm DM}}/{d\, \Omega} \rangle$ with the assumption that $\chi_2$ constitutes half of the total DM abundance~\cite{Krnjaic:2025zjl}.

\begin{figure}[t!]
    \centering
    \subfigure[\label{i1}]{
     \includegraphics[width=0.95\linewidth]{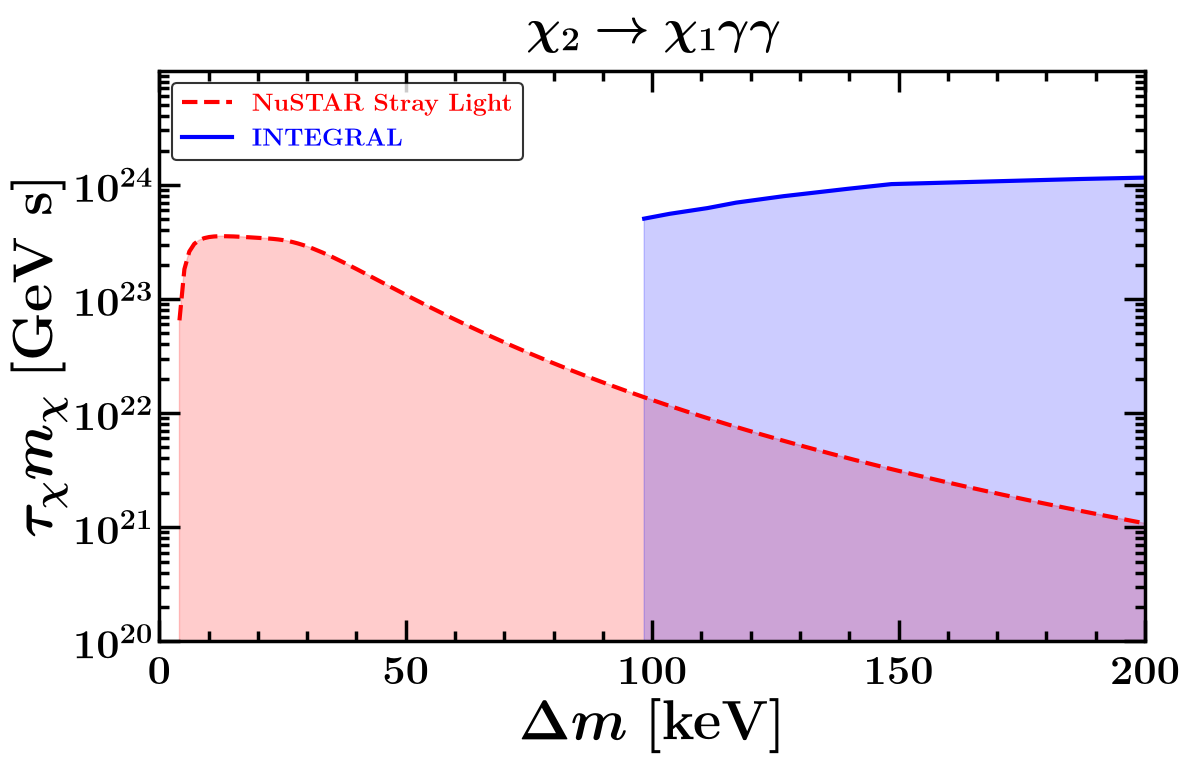}}
    \subfigure[\label{i2}]{
    \includegraphics[width=0.95\linewidth]{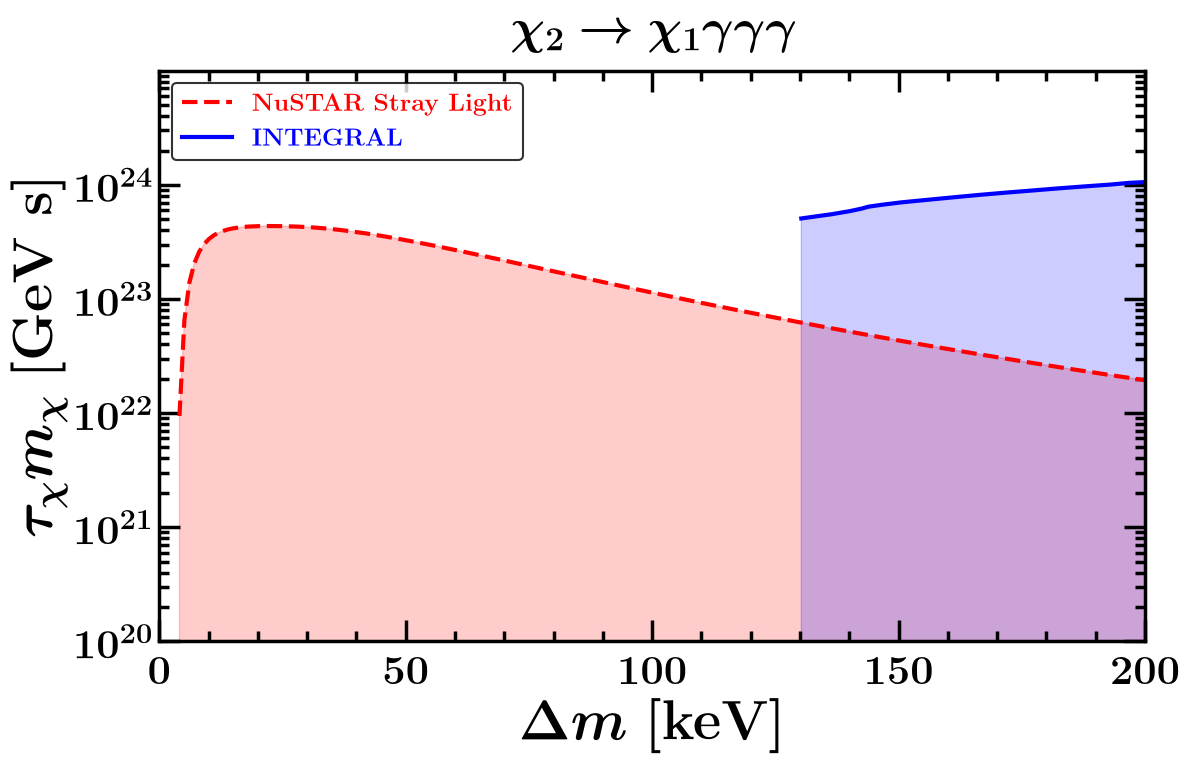}}
    \caption{Lower bound on the product of DM lifetime and mass ($\tau_\chi m_{\chi_2}$) from the NuSTAR SL data as function of mass splitting ($\Delta m$) (red dashed line) for (a) $\chi_2\to \chi_1+2 \gamma$ and (b) $\chi_2\to \chi_1+3 \gamma$.
    INTEGRAL \cite{Krnjaic:2025zjl} constraint is shown by the blue shaded region.}
    \label{fig:inela_DM}
\end{figure}

The photon spectrum for the inelastic DM decay is given by~\cite{Krnjaic:2025zjl}
\begin{equation}
    \dfrac{d \Phi_\gamma}{d E_\gamma} = \dfrac{1}{4\, \pi\, m_\chi\tau_{\chi_2}}\,\dfrac{1}{\Delta m}\,\dfrac{dN}{dx}\,\dfrac{D}{2}
\end{equation}
where 
\begin{align}
    \dfrac{dN}{dx} &= 
    \begin{cases}
        280\, x^3\,(1-x)^3;\qquad &\chi_2 \rightarrow \chi_1 +2\,\gamma \\
        \begin{aligned}
            \dfrac{143x^3}{1530}[(1-x) F(x)\\
              -11033820x^2 \log x];
        \end{aligned}
        \qquad &\chi_2 \rightarrow \chi_1 + 3\,\gamma 
    \end{cases}
\end{align}
with
\begin{align}
    F(x) =&\, 137200 - 2377850\,x - 15834839\,x^2 \nn \\
    &+ 14321161\,x^3 - 13688639\,x^4 \nn \\
    &+ 10465561\,x^5 - 5630414\,x^6 \nn \\
    &+ 1849450\,x^7 - 275450\,x^8
\end{align}
and $0 < x = E_\gamma/\Delta m < 1$. The coefficients of $dN/dx$ are set by the normalization condition $\int_0^1 dx(dN/dx) = 2\, (3)$ for the two (three) photon decay processes, respectively.

Using the photon spectrum given above, we calculate the expected photon flux from inelastic DM decay in the galaxy and the resulting spectra for $\chi_2\to \chi_1 + 2\,\gamma $ (red solid line) and $\chi_2\to \chi_1 + 3\,\gamma$ (blue solid line) for $\Delta m=20$ keV with $\tau_{\chi_2} m_\chi = 10^{20}$ are shown in Fig.\ref{fig:spec_inela}.
The observed NuSTAR SL spectra are also shown by the black dashed line.
Note that the two decay modes do not lead to a sharp photon spectrum, but rather a continuous distribution peaked at $\sim\Delta m/2$ for the two photon final state and peaked at $\sim\Delta m/3$ for the three photon final state.
This continuous distribution of the photons from the inelastic DM decay provides the scope for probing $\Delta m> 18$ keV.

Finally, following the procedure outlined in Sec.~\ref{sec:stat}, we derive the lower limit on the DM lifetime from the NuSTAR SL data, which are shown in Fig.~\ref{fig:inela_DM} in the $\Delta m$ vs.\ $\tau_\chi m_{\chi_2}$ plane by the red dashed line. The top and bottom panels correspond to the decay channels $\chi_2 \to \chi_1 + 2\gamma$ and $\chi_2 \to \chi_1 + 3\gamma$, respectively.

As discussed earlier, we find that the continuous photon spectrum from these decays enables NuSTAR to probe mass splittings as large as $\Delta m \gtrsim 30$ keV. However, for larger splittings, $\Delta m \gtrsim 100~(130)$ keV, the most stringent constraints on $\chi_2 \to \chi_1 + 2\gamma$ ($\chi_2 \to \chi_1 + 3\gamma$) are provided by INTEGRAL \cite{Krnjaic:2025zjl}, as indicated by the blue shaded region.
On the other hand, the lowest $\Delta m (\sim 3$ keV) one can probe using NuSTAR SL data is limited by the lowest energy bin available in the NuSTAR SL data.
Although for our analysis, we assumed the DM $\chi_2$ to satisfy half of the total observed DM density, one can also consider it to saturate the total DM density, and in that case, the corresponding limit can be obtained by appropriate scaling.

\section{Discussions}
\label{sec:discussions}
NuSTAR SL observations offer a unique window into diffuse X-ray emission, enabling sensitive searches for light dark matter in the $3$--$18$ keV energy range. In this work, we exploit this capability to hunt for spectral signatures of keV-scale decaying dark matter and derive constraints on a broad class of well-motivated particle physics models. For a given model, we calculate the predicted spectral signature of DM decay, and in the absence of any statistically significant excess in the diffuse X-ray spectrum, we derive  conservative upper bounds on the DM coupling to SM species.

For a scalar DM with Yukawa-like interaction with SM electrons, which can decay to two photons via an electron loop, we find that the NuSTAR SL data 
provides one of the strongest upper bounds on $e'_e$
for DM masses in the range $\sim 7 - 36$ keV.
For example, it probes the scalar DM coupling down to $ 4 \times 10^{-19}~ ( 6 \times 10^{-20})$ for DM mass 7 keV (36 keV).
For an ALP DM we consider both a photophilic ALP and an electrophilic ALP. We find that for the photophilic ALP DM, the NuSTAR SL data can constrain 
the ALP photon-coupling $\gagg \lesssim 10^{-18}\,\,\text{GeV}^{-1}$ ($\lesssim 2 \times 10^{-19}\,\,\text{GeV}^{-1}$) for $m_a\sim 7$ keV (36 keV) providing the strongest limit in the range $7 \lesssim m_a \lesssim 36$ keV.
On the other hand, for the electrophilic ALP DM, which can decay to two photons via an electron loop, the NuSTAR SL data can help to constrain the ALP-electron coupling over the same mass range as the photophilic case. 
NuSTAR SL data can constrain $\gae \lesssim 1.5 \times 10^{-14}~(8 \times 10^{-16})$ for $m_a\sim 7$ keV (36 keV).
Note that for both the scalar DM and the ALP DM, the primary decay channel is the decay to two photons either at tree level or at one-loop, leading to a monochromatic signal. On the contrary, for DP DM, the only decay channel to photons is the loop induced decay to three photons, also known as the \textit{trident decay}. Such a three-body decay results in a continuous photon spectrum as opposed to a monochromatic spectrum for the two-body case. This feature enables us to probe a wide range of DM masses using the NuSTAR SL data. Specifically, we find that for DP DM masses in the range $20$ keV to $60$ keV, using the NuSTAR SL data, one can get strongest indirect search limit on the DP-photon kinetic mixing.
For $m_{A'}\sim 22$ keV (65 keV) the upper limit on $\epsilon$ is found to be $\lesssim 2 \times 10^{-11}~(3 \times 10^{-13})$. Finally, we highlight the detection prospects of 
inelastic DM scenarios featuring multiple dark sector particles.
For such DM candidates, we 
can place an upper bound on the lifetime of the heavier DM particle decaying to $2\gamma$ and $3\gamma$ along with a lighter dark sector particle.
The energy of the signal photon is strongly correlated to the mass splitting $\Delta m$ between the dark sector particles.
Using NuSTAR stray light data, we place upper bounds on the DM lifetime 
for $3~{\rm keV} \lesssim \Delta m \lesssim 100~{\rm keV}$, which represent 
the strongest indirect detection constraints in this mass splitting range, superseding 
the previous bounds from INTEGRAL.
With $2\gamma$ signal NuSTAR SL data can probe $\tau_\chi m_\chi\lesssim 3\times 10^{23}$ GeV.s ($1.5 \times 10^{22}$ GeV.s) for $\Delta m \sim 10$ keV (100 keV).
On the other hand, with $3\gamma$ signal, it excludes $\tau_\chi m_\chi\lesssim 4\times 10^{23}$ GeV.s ($8 \times 10^{22}$ GeV.s) for $\Delta m \sim 10$ keV (130 keV).
Through this analysis, we highlight the fact that the NuSTAR SL data can be extremely valuable in astrophysical searches for keV-scale new physics.

Finally, we note that the bounds derived in this work adopt a conservative
treatment of the NuSTAR SL background, in which the full stacked diffuse
spectrum is treated as background without further decomposition.
Refs.~\cite{Krivonos:2024yvm,Zakharov:2025coj} have shown that this stacked
spectrum is well described by a combination of solar X-ray emission and the
cosmic X-ray background. 
A dedicated analysis involving astrophysical modeling of the background
and searching for a DM signal in the
residual might improve the obtained limits.

\section*{Acknowledgements}
SJ acknowledges financial support from the National Natural Science
Foundation of China (12425506 and 12375101) and State Key Laboratory of Dark Matter Physics.
TK acknowledges support in the form of Senior Research Fellowship from the Council of Scientific \& Industrial Research (CSIR), Government of India.

\bibliography{refs}

@article{Zwicky:1933gu,
    author = "Zwicky, F.",
    title = "{Die Rotverschiebung von extragalaktischen Nebeln}",
    doi = "10.1007/s10714-008-0707-4",
    journal = "Helv. Phys. Acta",
    volume = "6",
    pages = "110--127",
    year = "1933"
}

@article{Rubin:1970zza,
    author = "Rubin, Vera C. and Ford, Jr., W. Kent",
    title = "{Rotation of the Andromeda Nebula from a Spectroscopic Survey of Emission Regions}",
    doi = "10.1086/150317",
    journal = "Astrophys. J.",
    volume = "159",
    pages = "379--403",
    year = "1970"
}

@article{Clowe:2006eq,
    author = "Clowe, Douglas and Bradac, Marusa and Gonzalez, Anthony H. and Markevitch, Maxim and Randall, Scott W. and Jones, Christine and Zaritsky, Dennis",
    title = "{A direct empirical proof of the existence of dark matter}",
    eprint = "astro-ph/0608407",
    archivePrefix = "arXiv",
    reportNumber = "SLAC-PUB-12078",
    doi = "10.1086/508162",
    journal = "Astrophys. J. Lett.",
    volume = "648",
    pages = "L109--L113",
    year = "2006"
}

@article{Planck:2018vyg,
    author = "Aghanim, N. and others",
    collaboration = "Planck",
    title = "{Planck 2018 results. VI. Cosmological parameters}",
    eprint = "1807.06209",
    archivePrefix = "arXiv",
    primaryClass = "astro-ph.CO",
    doi = "10.1051/0004-6361/201833910",
    journal = "Astron. Astrophys.",
    volume = "641",
    pages = "A6",
    year = "2020",
    note = "[Erratum: Astron.Astrophys. 652, C4 (2021)]"
}

@article{Arcadi:2017kky,
    author = "Arcadi, Giorgio and Dutra, Ma\'\i{}ra and Ghosh, Pradipta and Lindner, Manfred and Mambrini, Yann and Pierre, Mathias and Profumo, Stefano and Queiroz, Farinaldo S.",
    title = "{The waning of the WIMP? A review of models, searches, and constraints}",
    eprint = "1703.07364",
    archivePrefix = "arXiv",
    primaryClass = "hep-ph",
    doi = "10.1140/epjc/s10052-018-5662-y",
    journal = "Eur. Phys. J. C",
    volume = "78",
    number = "3",
    pages = "203",
    year = "2018"
}

@article{Roszkowski:2017nbc,
    author = "Roszkowski, Leszek and Sessolo, Enrico Maria and Trojanowski, Sebastian",
    title = "{WIMP dark matter candidates and searches\textemdash{}current status and future prospects}",
    eprint = "1707.06277",
    archivePrefix = "arXiv",
    primaryClass = "hep-ph",
    reportNumber = "UCI-HEP-TR-2017-09, DO-TH-17-15, UCI-HEP-TR-2017-09-",
    doi = "10.1088/1361-6633/aab913",
    journal = "Rept. Prog. Phys.",
    volume = "81",
    number = "6",
    pages = "066201",
    year = "2018"
}

@article{Cirelli:2024ssz,
    author = "Cirelli, Marco and Strumia, Alessandro and Zupan, Jure",
    title = "{Dark Matter}",
    eprint = "2406.01705",
    archivePrefix = "arXiv",
    primaryClass = "hep-ph",
    month = "6",
    year = "2024",
    journal = ""
}

@article{ATLAS:2017bfj,
    author = "Aaboud, Morad and others",
    collaboration = "ATLAS",
    title = "{Search for dark matter and other new phenomena in events with an energetic jet and large missing transverse momentum using the ATLAS detector}",
    eprint = "1711.03301",
    archivePrefix = "arXiv",
    primaryClass = "hep-ex",
    reportNumber = "CERN-EP-2017-230",
    doi = "10.1007/JHEP01(2018)126",
    journal = "JHEP",
    volume = "01",
    pages = "126",
    year = "2018"
}

@article{CMS:2017zts,
    author = "Sirunyan, A. M. and others",
    collaboration = "CMS",
    title = "{Search for new physics in final states with an energetic jet or a hadronically decaying $W$ or $Z$ boson and transverse momentum imbalance at $\sqrt{s}=13\text{ }\text{ }\mathrm{TeV}$}",
    eprint = "1712.02345",
    archivePrefix = "arXiv",
    primaryClass = "hep-ex",
    reportNumber = "CMS-EXO-16-048, CERN-EP-2017-294",
    doi = "10.1103/PhysRevD.97.092005",
    journal = "Phys. Rev. D",
    volume = "97",
    number = "9",
    pages = "092005",
    year = "2018"
}

@article{Kahlhoefer:2017dnp,
    author = "Kahlhoefer, Felix",
    title = "{Review of LHC Dark Matter Searches}",
    eprint = "1702.02430",
    archivePrefix = "arXiv",
    primaryClass = "hep-ph",
    reportNumber = "DESY-17-024",
    doi = "10.1142/S0217751X1730006X",
    journal = "Int. J. Mod. Phys. A",
    volume = "32",
    number = "13",
    pages = "1730006",
    year = "2017"
}

@article{Gaskins:2016cha,
    author = "Gaskins, Jennifer M.",
    title = "{A review of indirect searches for particle dark matter}",
    eprint = "1604.00014",
    archivePrefix = "arXiv",
    primaryClass = "astro-ph.HE",
    doi = "10.1080/00107514.2016.1175160",
    journal = "Contemp. Phys.",
    volume = "57",
    number = "4",
    pages = "496--525",
    year = "2016"
}

@article{PerezdelosHeros:2020qyt,
    author = "P{\'e}rez de los Heros, Carlos",
    title = "{Status, Challenges and Directions in Indirect Dark Matter Searches}",
    eprint = "2008.11561",
    archivePrefix = "arXiv",
    primaryClass = "astro-ph.HE",
    doi = "10.3390/sym12101648",
    journal = "Symmetry",
    volume = "12",
    number = "10",
    pages = "1648",
    year = "2020"
}

@article{Misiaszek:2023sxe,
    author = "Misiaszek, Marcin and Rossi, Nicola",
    title = "{Direct Detection of Dark Matter: A Critical Review}",
    eprint = "2310.20472",
    archivePrefix = "arXiv",
    primaryClass = "hep-ph",
    doi = "10.3390/sym16020201",
    journal = "Symmetry",
    volume = "16",
    number = "2",
    pages = "201",
    year = "2024"
}

@article{Billard:2021uyg,
    author = "Billard, Julien and others",
    title = "{Direct detection of dark matter{\textemdash}APPEC committee report*}",
    eprint = "2104.07634",
    archivePrefix = "arXiv",
    primaryClass = "hep-ex",
    doi = "10.1088/1361-6633/ac5754",
    journal = "Rept. Prog. Phys.",
    volume = "85",
    number = "5",
    pages = "056201",
    year = "2022"
}

@article{Scherrer:1985zt,
    author = "Scherrer, Robert J. and Turner, Michael S.",
    title = "{On the Relic, Cosmic Abundance of Stable Weakly Interacting Massive Particles}",
    reportNumber = "FERMILAB-PUB-85-163-A, EFI-85-76-CHICAGO",
    doi = "10.1103/PhysRevD.33.1585",
    journal = "Phys. Rev. D",
    volume = "33",
    pages = "1585",
    year = "1986",
    note = "[Erratum: Phys.Rev.D 34, 3263 (1986)]"
}

@article{Srednicki:1988ce,
    author = "Srednicki, Mark and Watkins, Richard and Olive, Keith A.",
    editor = "Srednicki, M. A.",
    title = "{Calculations of Relic Densities in the Early Universe}",
    reportNumber = "UMN-TH-646/88",
    doi = "10.1016/0550-3213(88)90099-5",
    journal = "Nucl. Phys. B",
    volume = "310",
    pages = "693",
    year = "1988"
}

@article{Krivonos:2024yvm,
    author = "Krivonos, R. A. and Barinov, V. V. and Mukhin, A. A. and Gorbunov, D. S.",
    title = "{Strong Limits on keV-Scale Galactic Sterile Neutrino Dark Matter with Stray Light from NuSTAR after 11~Years of Operation}",
    eprint = "2405.17861",
    archivePrefix = "arXiv",
    primaryClass = "hep-ph",
    doi = "10.1103/PhysRevLett.133.261002",
    journal = "Phys. Rev. Lett.",
    volume = "133",
    number = "26",
    pages = "261002",
    year = "2024"
}

@article{Zakharov:2025coj,
    author = "Zakharov, E. I. and Barinov, V. V. and Gorbunov, D. S. and Krivonos, R. A. and Mukhin, A. A.",
    title = "{Search for a photon peak from keV-scale dark matter annihilation with NuSTAR: Constraints on {\ensuremath{\langle}}{\ensuremath{\sigma}}v{\ensuremath{\rangle}} after 11~years of observations}",
    eprint = "2509.08506",
    archivePrefix = "arXiv",
    primaryClass = "astro-ph.HE",
    reportNumber = "INR-TH-2025-013",
    doi = "10.1103/f1w8-bcny",
    journal = "Phys. Rev. D",
    volume = "112",
    number = "10",
    pages = "103037",
    year = "2025"
}

@article{Ferreira:2022egk,
    author = {Ferreira, Ricardo Z. and Marsh, M. C. David and M{\"u}ller, Eike},
    title = "{Do Direct Detection Experiments Constrain Axionlike Particles Coupled to Electrons?}",
    eprint = "2202.08858",
    archivePrefix = "arXiv",
    primaryClass = "hep-ph",
    doi = "10.1103/PhysRevLett.128.221302",
    journal = "Phys. Rev. Lett.",
    volume = "128",
    number = "22",
    pages = "221302",
    year = "2022"
}

@article{Linden:2024fby,
    author = "Linden, Tim and Nguyen, Thong T. Q. and Tait, Tim M. P.",
    title = "{X-ray constraints on dark photon tridents}",
    eprint = "2406.19445",
    archivePrefix = "arXiv",
    primaryClass = "hep-ph",
    reportNumber = "UCI-HEP-TR-2024-10",
    doi = "10.1103/37gn-x3y1",
    journal = "Phys. Rev. D",
    volume = "112",
    number = "2",
    pages = "023026",
    year = "2025"
}

@article{Bickendorf:2022buy,
    author = "Bickendorf, Gerrit and Drees, Manuel",
    title = "{Constraints on light leptophilic dark matter mediators from decay experiments}",
    eprint = "2206.05038",
    archivePrefix = "arXiv",
    primaryClass = "hep-ph",
    doi = "10.1140/epjc/s10052-022-11128-9",
    journal = "Eur. Phys. J. C",
    volume = "82",
    number = "12",
    pages = "1163",
    year = "2022"
}

@article{Bauer:2017ris,
    author = "Bauer, Martin and Neubert, Matthias and Thamm, Andrea",
    title = "{Collider Probes of Axion-Like Particles}",
    eprint = "1708.00443",
    archivePrefix = "arXiv",
    primaryClass = "hep-ph",
    reportNumber = "MITP-17-047",
    doi = "10.1007/JHEP12(2017)044",
    journal = "JHEP",
    volume = "12",
    pages = "044",
    year = "2017"
}

@article{Montefalcone:2025nmm,
    author = "Montefalcone, Gabriele and Elor, Gilly and Boddy, Kimberly K. and Bellomo, Nicola",
    title = "{CMB constraints on loop-induced decays of leptophilic dark matter}",
    eprint = "2503.00110",
    archivePrefix = "arXiv",
    primaryClass = "hep-ph",
    reportNumber = "UTWI-02-2025",
    doi = "10.1103/hxmm-vcjz",
    journal = "Phys. Rev. D",
    volume = "112",
    number = "2",
    pages = "023506",
    year = "2025"
}

@article{Cowan:2010js,
    author = "Cowan, Glen and Cranmer, Kyle and Gross, Eilam and Vitells, Ofer",
    title = "{Asymptotic formulae for likelihood-based tests of new physics}",
    eprint = "1007.1727",
    archivePrefix = "arXiv",
    primaryClass = "physics.data-an",
    doi = "10.1140/epjc/s10052-011-1554-0",
    journal = "Eur. Phys. J. C",
    volume = "71",
    pages = "1554",
    year = "2011",
    note = "[Erratum: Eur.Phys.J.C 73, 2501 (2013)]"
}

@article{Knapen:2017xzo,
    author = "Knapen, Simon and Lin, Tongyan and Zurek, Kathryn M.",
    title = "{Light Dark Matter: Models and Constraints}",
    eprint = "1709.07882",
    archivePrefix = "arXiv",
    primaryClass = "hep-ph",
    doi = "10.1103/PhysRevD.96.115021",
    journal = "Phys. Rev. D",
    volume = "96",
    number = "11",
    pages = "115021",
    year = "2017"
}

@article{Mitridate:2021ctr,
    author = "Mitridate, Andrea and Trickle, Tanner and Zhang, Zhengkang and Zurek, Kathryn M.",
    title = "{Dark matter absorption via electronic excitations}",
    eprint = "2106.12586",
    archivePrefix = "arXiv",
    primaryClass = "hep-ph",
    reportNumber = "CALT-TH-2021-025",
    doi = "10.1007/JHEP09(2021)123",
    journal = "JHEP",
    volume = "09",
    pages = "123",
    year = "2021"
}

@article{Batell:2016ove,
    author = "Batell, Brian and Lange, Nicholas and McKeen, David and Pospelov, Maxim and Ritz, Adam",
    title = "{Muon anomalous magnetic moment through the leptonic Higgs portal}",
    eprint = "1606.04943",
    archivePrefix = "arXiv",
    primaryClass = "hep-ph",
    doi = "10.1103/PhysRevD.95.075003",
    journal = "Phys. Rev. D",
    volume = "95",
    number = "7",
    pages = "075003",
    year = "2017"
}

@article{Foster:2021ngm,
    author = "Foster, Joshua W. and Kongsore, Marius and Dessert, Christopher and Park, Yujin and Rodd, Nicholas L. and Cranmer, Kyle and Safdi, Benjamin R.",
    title = "{Deep Search for Decaying Dark Matter with XMM-Newton Blank-Sky Observations}",
    eprint = "2102.02207",
    archivePrefix = "arXiv",
    primaryClass = "astro-ph.CO",
    reportNumber = "LCTP-21-05",
    doi = "10.1103/PhysRevLett.127.051101",
    journal = "Phys. Rev. Lett.",
    volume = "127",
    number = "5",
    pages = "051101",
    year = "2021"
}

@article{Ng:2019gch,
    author = "Ng, Kenny C. Y. and Roach, Brandon M. and Perez, Kerstin and Beacom, John F. and Horiuchi, Shunsaku and Krivonos, Roman and Wik, Daniel R.",
    title = "{New Constraints on Sterile Neutrino Dark Matter from $NuSTAR$ M31 Observations}",
    eprint = "1901.01262",
    archivePrefix = "arXiv",
    primaryClass = "astro-ph.HE",
    doi = "10.1103/PhysRevD.99.083005",
    journal = "Phys. Rev. D",
    volume = "99",
    pages = "083005",
    year = "2019"
}

@article{Hardy:2016kme,
    author = "Hardy, Edward and Lasenby, Robert",
    title = "{Stellar cooling bounds on new light particles: plasma mixing effects}",
    eprint = "1611.05852",
    archivePrefix = "arXiv",
    primaryClass = "hep-ph",
    doi = "10.1007/JHEP02(2017)033",
    journal = "JHEP",
    volume = "02",
    pages = "033",
    year = "2017"
}

@article{Laha:2020ivk,
    author = "Laha, Ranjan and Mu{\~n}oz, Julian B. and Slatyer, Tracy R.",
    title = "{INTEGRAL constraints on primordial black holes and particle dark matter}",
    eprint = "2004.00627",
    archivePrefix = "arXiv",
    primaryClass = "astro-ph.CO",
    doi = "10.1103/PhysRevD.101.123514",
    journal = "Phys. Rev. D",
    volume = "101",
    number = "12",
    pages = "123514",
    year = "2020"
}

@article{Preskill:1982cy,
    author = "Preskill, John and Wise, Mark B. and Wilczek, Frank",
    editor = "Srednicki, M. A.",
    title = "{Cosmology of the Invisible Axion}",
    reportNumber = "HUTP-82-A048, NSF-ITP-82-103",
    doi = "10.1016/0370-2693(83)90637-8",
    journal = "Phys. Lett. B",
    volume = "120",
    pages = "127--132",
    year = "1983"
}

@article{Dine:1982ah,
    author = "Dine, Michael and Fischler, Willy",
    editor = "Srednicki, M. A.",
    title = "{The Not So Harmless Axion}",
    reportNumber = "UPR-0201T",
    doi = "10.1016/0370-2693(83)90639-1",
    journal = "Phys. Lett. B",
    volume = "120",
    pages = "137--141",
    year = "1983"
}

@article{Abbott:1982af,
    author = "Abbott, L. F. and Sikivie, P.",
    editor = "Srednicki, M. A.",
    title = "{A Cosmological Bound on the Invisible Axion}",
    reportNumber = "PRINT-82-0695 (BRANDEIS)",
    doi = "10.1016/0370-2693(83)90638-X",
    journal = "Phys. Lett. B",
    volume = "120",
    pages = "133--136",
    year = "1983"
}

@article{DiLuzio:2020wdo,
    author = "Di Luzio, Luca and Giannotti, Maurizio and Nardi, Enrico and Visinelli, Luca",
    title = "{The landscape of QCD axion models}",
    eprint = "2003.01100",
    archivePrefix = "arXiv",
    primaryClass = "hep-ph",
    reportNumber = "DESY 20-036, DESY-20-036",
    doi = "10.1016/j.physrep.2020.06.002",
    journal = "Phys. Rept.",
    volume = "870",
    pages = "1--117",
    year = "2020"
}

@article{Co:2019jts,
    author = "Co, Raymond T. and Hall, Lawrence J. and Harigaya, Keisuke",
    title = "{Axion Kinetic Misalignment Mechanism}",
    eprint = "1910.14152",
    archivePrefix = "arXiv",
    primaryClass = "hep-ph",
    reportNumber = "LCTP-19-28",
    doi = "10.1103/PhysRevLett.124.251802",
    journal = "Phys. Rev. Lett.",
    volume = "124",
    number = "25",
    pages = "251802",
    year = "2020"
}

@article{Weinberg:1977ma,
    author = "Weinberg, Steven",
    title = "{A New Light Boson?}",
    reportNumber = "HUTP-77/A074",
    doi = "10.1103/PhysRevLett.40.223",
    journal = "Phys. Rev. Lett.",
    volume = "40",
    pages = "223--226",
    year = "1978"
}

@article{Wilczek:1977pj,
    author = "Wilczek, Frank",
    title = "{Problem of Strong  $P$  and  $T$  Invariance in the Presence of Instantons}",
    reportNumber = "Print-77-0939 (COLUMBIA)",
    doi = "10.1103/PhysRevLett.40.279",
    journal = "Phys. Rev. Lett.",
    volume = "40",
    pages = "279--282",
    year = "1978"
}

@article{Peccei:1977hh,
    author = "Peccei, R. D. and Quinn, Helen R.",
    title = "{CP Conservation in the Presence of Instantons}",
    reportNumber = "ITP-568-STANFORD",
    doi = "10.1103/PhysRevLett.38.1440",
    journal = "Phys. Rev. Lett.",
    volume = "38",
    pages = "1440--1443",
    year = "1977"
}

@article{Kim:2008hd,
    author = "Kim, Jihn E. and Carosi, Gianpaolo",
    title = "{Axions and the Strong CP Problem}",
    eprint = "0807.3125",
    archivePrefix = "arXiv",
    primaryClass = "hep-ph",
    doi = "10.1103/RevModPhys.82.557",
    journal = "Rev. Mod. Phys.",
    volume = "82",
    pages = "557--602",
    year = "2010",
    note = "[Erratum: Rev.Mod.Phys. 91, 049902 (2019)]"
}

@inbook{Biekotter:2025fll,
    author = {Biek{\"o}tter, Anke and Mimasu, Ken},
    title = "{Axions and Axion-like particles: collider searches}",
    eprint = "2508.19358",
    archivePrefix = "arXiv",
    primaryClass = "hep-ph",
    month = "8",
    year = "2025"
}

@article{Bharucha:2022lty,
    author = {Bharucha, A. and Br\"ummer, F. and Desai, N. and Mutzel, S.},
    title = "{Axion-like particles as mediators for dark matter: beyond freeze-out}",
    eprint = "2209.03932",
    archivePrefix = "arXiv",
    primaryClass = "hep-ph",
    doi = "10.1007/JHEP02(2023)141",
    journal = "JHEP",
    volume = "02",
    pages = "141",
    year = "2023"
}

@article{Pospelov:2008jk,
    author = "Pospelov, Maxim and Ritz, Adam and Voloshin, Mikhail B.",
    title = "{Bosonic super-WIMPs as keV-scale dark matter}",
    eprint = "0807.3279",
    archivePrefix = "arXiv",
    primaryClass = "hep-ph",
    reportNumber = "FTPI-MINN-08-29, UMN-TH-2709-08",
    doi = "10.1103/PhysRevD.78.115012",
    journal = "Phys. Rev. D",
    volume = "78",
    pages = "115012",
    year = "2008"
}

@article{Nelson:2011sf,
    author = "Nelson, Ann E. and Scholtz, Jakub",
    title = "{Dark Light, Dark Matter and the Misalignment Mechanism}",
    eprint = "1105.2812",
    archivePrefix = "arXiv",
    primaryClass = "hep-ph",
    doi = "10.1103/PhysRevD.84.103501",
    journal = "Phys. Rev. D",
    volume = "84",
    pages = "103501",
    year = "2011"
}

@article{Caputo:2021eaa,
    author = "Caputo, Andrea and Millar, Alexander J. and O'Hare, Ciaran A. J. and Vitagliano, Edoardo",
    title = "{Dark photon limits: A handbook}",
    eprint = "2105.04565",
    archivePrefix = "arXiv",
    primaryClass = "hep-ph",
    reportNumber = "NORDITA-2021-036",
    doi = "10.1103/PhysRevD.104.095029",
    journal = "Phys. Rev. D",
    volume = "104",
    number = "9",
    pages = "095029",
    year = "2021"
}

@article{Redondo:2008ec,
    author = "Redondo, Javier and Postma, Marieke",
    title = "{Massive hidden photons as lukewarm dark matter}",
    eprint = "0811.0326",
    archivePrefix = "arXiv",
    primaryClass = "hep-ph",
    reportNumber = "NIKHEF-2008-030, DESY-08-154",
    doi = "10.1088/1475-7516/2009/02/005",
    journal = "JCAP",
    volume = "02",
    pages = "005",
    year = "2009"
}

@article{Arias:2012az,
    author = "Arias, Paola and Cadamuro, Davide and Goodsell, Mark and Jaeckel, Joerg and Redondo, Javier and Ringwald, Andreas",
    title = "{WISPy Cold Dark Matter}",
    eprint = "1201.5902",
    archivePrefix = "arXiv",
    primaryClass = "hep-ph",
    reportNumber = "DESY-11-226, MPP-2011-140, CERN-PH-TH-2011-323, IPPP-11-80, DCPT-11-160",
    doi = "10.1088/1475-7516/2012/06/013",
    journal = "JCAP",
    volume = "06",
    pages = "013",
    year = "2012"
}

@article{Rizzo:2018ntg,
    author = "Rizzo, Thomas G.",
    title = "{Kinetic mixing, dark photons and an extra dimension. Part I}",
    eprint = "1801.08525",
    archivePrefix = "arXiv",
    primaryClass = "hep-ph",
    reportNumber = "SLAC-PUB-17178",
    doi = "10.1007/JHEP07(2018)118",
    journal = "JHEP",
    volume = "07",
    pages = "118",
    year = "2018"
}

@article{Landau:1948kw,
    author = "Landau, L. D.",
    title = "{On the angular momentum of a system of two photons}",
    doi = "10.1016/B978-0-08-010586-4.50070-5",
    journal = "Dokl. Akad. Nauk SSSR",
    volume = "60",
    number = "2",
    pages = "207--209",
    year = "1948"
}

@article{Yang:1950rg,
    author = "Yang, Chen-Ning",
    title = "{Selection Rules for the Dematerialization of a Particle Into Two Photons}",
    doi = "10.1103/PhysRev.77.242",
    journal = "Phys. Rev.",
    volume = "77",
    pages = "242--245",
    year = "1950"
}

@article{An:2013yfc,
    author = "An, Haipeng and Pospelov, Maxim and Pradler, Josef",
    title = "{New stellar constraints on dark photons}",
    eprint = "1302.3884",
    archivePrefix = "arXiv",
    primaryClass = "hep-ph",
    reportNumber = "PI-PARTPHYS-318",
    doi = "10.1016/j.physletb.2013.07.008",
    journal = "Phys. Lett. B",
    volume = "725",
    pages = "190--195",
    year = "2013"
}

@article{Giannotti:2015kwo,
    author = "Giannotti, Maurizio and Irastorza, Igor and Redondo, Javier and Ringwald, Andreas",
    title = "{Cool WISPs for stellar cooling excesses}",
    eprint = "1512.08108",
    archivePrefix = "arXiv",
    primaryClass = "astro-ph.HE",
    reportNumber = "DESY-15-245",
    doi = "10.1088/1475-7516/2016/05/057",
    journal = "JCAP",
    volume = "05",
    pages = "057",
    year = "2016"
}

@article{XENON:2021qze,
    author = "Aprile, E. and others",
    collaboration = "XENON",
    title = "{Emission of single and few electrons in XENON1T and limits on light dark matter}",
    eprint = "2112.12116",
    archivePrefix = "arXiv",
    primaryClass = "hep-ex",
    doi = "10.1103/PhysRevD.106.022001",
    journal = "Phys. Rev. D",
    volume = "106",
    number = "2",
    pages = "022001",
    year = "2022",
    note = "[Erratum: Phys.Rev.D 110, 109903 (2024)]"
}

@article{Krnjaic:2025zjl,
    author = "Krnjaic, Gordan and McKeen, David and Mizuta, Riku and Mohlabeng, Gopolang and Morrissey, David E. and Tuckler, Douglas",
    title = "{X-rays from inelastic dark matter freeze-in}",
    eprint = "2509.19428",
    archivePrefix = "arXiv",
    primaryClass = "hep-ph",
    reportNumber = "FERMILAB-PUB-25-0673-T",
    doi = "10.1103/99z7-kz4s",
    journal = "Phys. Rev. D",
    volume = "112",
    number = "11",
    pages = "115039",
    year = "2025"
}

@article{XENON:2020rca,
    author = "Aprile, E. and others",
    collaboration = "XENON",
    title = "{Excess electronic recoil events in XENON1T}",
    eprint = "2006.09721",
    archivePrefix = "arXiv",
    primaryClass = "hep-ex",
    doi = "10.1103/PhysRevD.102.072004",
    journal = "Phys. Rev. D",
    volume = "102",
    number = "7",
    pages = "072004",
    year = "2020"
}

@article{XENON:2022ltv,
    author = "Aprile, E. and others",
    collaboration = "XENON",
    title = "{Search for New Physics in Electronic Recoil Data from XENONnT}",
    eprint = "2207.11330",
    archivePrefix = "arXiv",
    primaryClass = "hep-ex",
    doi = "10.1103/PhysRevLett.129.161805",
    journal = "Phys. Rev. Lett.",
    volume = "129",
    number = "16",
    pages = "161805",
    year = "2022"
}

@article{Cadamuro:2011fd,
    author = "Cadamuro, Davide and Redondo, Javier",
    title = "{Cosmological bounds on pseudo Nambu-Goldstone bosons}",
    eprint = "1110.2895",
    archivePrefix = "arXiv",
    primaryClass = "hep-ph",
    reportNumber = "MPP-2011-116",
    doi = "10.1088/1475-7516/2012/02/032",
    journal = "JCAP",
    volume = "02",
    pages = "032",
    year = "2012"
}

@article{Cautun:2019eaf,
    author = "Cautun, Marius and Benitez-Llambay, Alejandro and Deason, Alis J. and Frenk, Carlos S. and Fattahi, Azadeh and G{\'o}mez, Facundo A. and Grand, Robert J. J. and Oman, Kyle A. and Navarro, Julio F. and Simpson, Christine M.",
    title = "{The Milky Way total mass profile as inferred from Gaia DR2}",
    eprint = "1911.04557",
    archivePrefix = "arXiv",
    primaryClass = "astro-ph.GA",
    doi = "10.1093/mnras/staa1017",
    journal = "Mon. Not. Roy. Astron. Soc.",
    volume = "494",
    number = "3",
    pages = "4291--4313",
    year = "2020"
}

@misc{NuSTAR_tech_desc,
  author       = {{NASA HEASARC}},
  title        = {NuSTAR Technical Description},
  year         = {2023},
  howpublished = {\url{https://heasarc.gsfc.nasa.gov/docs/nustar/nustar_tech_desc.html}},
  note         = {Accessed: 2026-04-19}
}

@article{Fermi-LAT:2016afa,
    author = "Charles, E. and others",
    collaboration = "Fermi-LAT",
    title = "{Sensitivity Projections for Dark Matter Searches with the Fermi Large Area Telescope}",
    eprint = "1605.02016",
    archivePrefix = "arXiv",
    primaryClass = "astro-ph.HE",
    reportNumber = "FERMILAB-PUB-16-179-AE",
    doi = "10.1016/j.physrep.2016.05.001",
    journal = "Phys. Rept.",
    volume = "636",
    pages = "1--46",
    year = "2016"
}

@article{Bergstrom:2013jra,
    author = "Bergstrom, Lars and Bringmann, Torsten and Cholis, Ilias and Hooper, Dan and Weniger, Christoph",
    title = "{New Limits on Dark Matter Annihilation from AMS Cosmic Ray Positron Data}",
    eprint = "1306.3983",
    archivePrefix = "arXiv",
    primaryClass = "astro-ph.HE",
    reportNumber = "FERMILAB-PUB-13-202-A",
    doi = "10.1103/PhysRevLett.111.171101",
    journal = "Phys. Rev. Lett.",
    volume = "111",
    pages = "171101",
    year = "2013"
}

@article{AMS:2019iwo,
    author = "Aguilar, M. and others",
    collaboration = "AMS",
    title = "{Towards Understanding the Origin of Cosmic-Ray Electrons}",
    doi = "10.1103/PhysRevLett.122.101101",
    journal = "Phys. Rev. Lett.",
    volume = "122",
    number = "10",
    pages = "101101",
    year = "2019"
}

@article{HESS:2022ygk,
    author = "Abdalla, H. and others",
    collaboration = "H.E.S.S.",
    title = "{Search for Dark Matter Annihilation Signals in the H.E.S.S. Inner Galaxy Survey}",
    eprint = "2207.10471",
    archivePrefix = "arXiv",
    primaryClass = "astro-ph.HE",
    doi = "10.1103/PhysRevLett.129.111101",
    journal = "Phys. Rev. Lett.",
    volume = "129",
    number = "11",
    pages = "111101",
    year = "2022"
}

@article{Baur:2019jwm,
    author = "Baur, Sebastian",
    collaboration = "IceCube",
    title = "{Dark matter searches with the IceCube Upgrade}",
    eprint = "1908.08236",
    archivePrefix = "arXiv",
    primaryClass = "astro-ph.HE",
    reportNumber = "PoS-ICRC2019-506",
    doi = "10.22323/1.358.0506",
    journal = "PoS",
    volume = "ICRC2019",
    pages = "506",
    year = "2020"
}

@article{Arguelles:2019ouk,
    author = {Arg{\"u}elles, Carlos A. and Diaz, Alejandro and Kheirandish, Ali and Olivares-Del-Campo, Andr{\'e}s and Safa, Ibrahim and Vincent, Aaron C.},
    title = "{Dark matter annihilation to neutrinos}",
    eprint = "1912.09486",
    archivePrefix = "arXiv",
    primaryClass = "hep-ph",
    doi = "10.1103/RevModPhys.93.035007",
    journal = "Rev. Mod. Phys.",
    volume = "93",
    number = "3",
    pages = "035007",
    year = "2021"
}

@article{Cirelli:2020bpc,
    author = "Cirelli, Marco and Fornengo, Nicolao and Kavanagh, Bradley J. and Pinetti, Elena",
    title = "{Integral X-ray constraints on sub-GeV Dark Matter}",
    eprint = "2007.11493",
    archivePrefix = "arXiv",
    primaryClass = "hep-ph",
    doi = "10.1103/PhysRevD.103.063022",
    journal = "Phys. Rev. D",
    volume = "103",
    number = "6",
    pages = "063022",
    year = "2021"
}

@article{Bouchet:2011fn,
    author = "Bouchet, Laurent and Strong, Andrew W. and Porter, Troy A. and Moskalenko, Igor V. and Jourdain, Elisabeth and Roques, Jean-Pierre",
    title = "{Diffuse emission measurement with INTEGRAL/SPI as indirect probe of cosmic-ray electrons and positrons}",
    eprint = "1107.0200",
    archivePrefix = "arXiv",
    primaryClass = "astro-ph.HE",
    doi = "10.1088/0004-637X/739/1/29",
    journal = "Astrophys. J.",
    volume = "739",
    pages = "29",
    year = "2011"
}

@article{Super-Kamiokande:2015qek,
    author = "Richard, E. and others",
    collaboration = "Super-Kamiokande",
    title = "{Measurements of the atmospheric neutrino flux by Super-Kamiokande: energy spectra, geomagnetic effects, and solar modulation}",
    eprint = "1510.08127",
    archivePrefix = "arXiv",
    primaryClass = "hep-ex",
    doi = "10.1103/PhysRevD.94.052001",
    journal = "Phys. Rev. D",
    volume = "94",
    number = "5",
    pages = "052001",
    year = "2016"
}

@article{COMPTEL,
       author = "Schoenfelder, V. and others",
        title = "{Instrument Description and Performance of the Imaging Gamma-Ray Telescope COMPTEL aboard the Compton Gamma-Ray Observatory}",
      journal = {The Astrophysical Journal Supplement Series},
         year = 1993,
        month = jun,
       volume = {86},
        pages = {657},
          doi = {10.1086/191794},
       adsurl = {https://ui.adsabs.harvard.edu/abs/1993ApJS...86..657S}
}

@article{Farzan:2014foo,
    author = "Farzan, Yasaman and Akbarieh, Amin Rezaei",
    title = "{Decaying Vector Dark Matter as an Explanation for the 3.5 keV Line from Galaxy Clusters}",
    eprint = "1408.2950",
    archivePrefix = "arXiv",
    primaryClass = "hep-ph",
    doi = "10.1088/1475-7516/2014/11/015",
    journal = "JCAP",
    volume = "11",
    pages = "015",
    year = "2014"
}

@article{DelaTorreLuque:2025zjt,
    author = "De la Torre Luque, Pedro and Carenza, Pierluca and Nguyen, Thong T. Q.",
    title = "{Sub-keV dark matter can strongly ionize molecular clouds}",
    eprint = "2507.01962",
    archivePrefix = "arXiv",
    primaryClass = "hep-ph",
    doi = "10.1103/qmw9-1k4k",
    journal = "Phys. Rev. D",
    volume = "113",
    number = "6",
    pages = "063035",
    year = "2026"
}

@article{Nguyen:2024kwy,
    author = "Nguyen, Thong T. Q. and John, Isabelle and Linden, Tim and Tait, Tim M. P.",
    title = "{Strong constraints on dark photon and scalar dark matter decay from INTEGRAL and AMS-02 data}",
    eprint = "2412.00180",
    archivePrefix = "arXiv",
    primaryClass = "hep-ph",
    reportNumber = "UCI-HEP-TR-2024-19",
    doi = "10.1103/5wrw-7dz8",
    journal = "Phys. Rev. D",
    volume = "113",
    number = "10",
    pages = "103051",
    year = "2026"
}

@article{Nguyen:2025tkl,
    author = "Nguyen, Thong T. Q. and De la Torre Luque, Pedro and John, Isabelle and Balaji, Shyam and Carenza, Pierluca and Linden, Tim",
    title = "{INTEGRAL, eROSITA and Voyager constraints on light bosonic dark matter: ALPs, dark photons, scalars, B-L and Li-Lj vectors}",
    eprint = "2507.13432",
    archivePrefix = "arXiv",
    primaryClass = "hep-ph",
    reportNumber = "KCL-2025-30",
    doi = "10.1103/fn6k-1nlc",
    journal = "Phys. Rev. D",
    volume = "113",
    number = "10",
    pages = "103010",
    year = "2026"
}

@article{TerolCalvo:2026nlr,
    author = "Terol Calvo, Jorge and Taoso, Marco and Caputo, Andrea and Negro, Michela and Regis, Marco",
    title = "{Searching for dark matter X-ray lines from the Large Magellanic Cloud with eROSITA}",
    eprint = "2603.19109",
    archivePrefix = "arXiv",
    primaryClass = "hep-ph",
    month=mar,
    year = "2026",
    journal=""
}

\end{document}